\documentclass[10pt,conference]{IEEEtran}
\usepackage{cite}
\usepackage{amsmath,amssymb,amsfonts}
\usepackage{graphicx}
\usepackage{textcomp}
\usepackage{xcolor}
\usepackage[hyphens]{url}
\usepackage{fancyhdr}
\usepackage{hyperref}
\usepackage{amsmath,amssymb,amsfonts}
\usepackage[c]{esvect}
\usepackage [ruled,linesnumbered] {algorithm2e} 
\usepackage[noend]{algorithmic}
\usepackage{cleveref}
\usepackage{subfig}
\usepackage{makecell}

\usepackage{multicol}
\usepackage{multirow}
\usepackage{tabularx}
\newcolumntype{C}[1]{>{\centering\arraybackslash}m{#1}}
\newcolumntype{R}[1]{>{\raggedleft\arraybackslash}m{#1}}
\newcolumntype{L}[1]{>{\raggedright\arraybackslash}m{#1}}
\usepackage{multirow}

\usepackage{tikz}
\newcommand*\cycle[1]{\tikz[baseline=(char.base)]{
            \node[shape=circle,fill,color=black,text=white,inner sep=0.05pt](char){#1};}}

\newcommand{\squishlist}{
  \begin{list}{$\bullet$}{
    \setlength{\itemsep}{0pt}       \setlength{\parsep}{3pt}
    \setlength{\topsep}{3pt}        \setlength{\partopsep}{0pt}
    \setlength{\leftmargin}{1em}    \setlength{\labelwidth}{1em}
    \setlength{\labelsep}{0.5em} } }

\newcommand{\squishend}{
  \end{list} }

\makeatletter 
\newcommand\notsotiny{\@setfontsize\tiny\@vipt\@viipt}
\makeatother
\usepackage{listings}
\definecolor{dkgreen}{rgb}{0,0.6,0}
\definecolor{gray}{rgb}{0.5,0.5,0.5}
\definecolor{mauve}{rgb}{0.58,0,0.82}
\definecolor{codeyellow}{rgb}{1.0, 0.49, 0.0}
\newcommand{\sysname}{{RPCAcc}}

\newcommand{\hpcayear}{2025}

\definecolor{colorRed}{RGB}{191,0,0}
\definecolor{colorGreen}{RGB}{0,170,62}
\definecolor{colorBlue}{RGB}{0,0,235}

\newcommand{\cBlue}[1]{{#1}}

\title{\sysname: A High-Performance and Reconfigurable PCIe-attached RPC Accelerator}

 \author{
 Jie Zhang$^{\dag,1}$,  
 Hongjing Huang$^{\dag,1}$,  
 Xuzheng Xu$^{1}$,
 Xiang Li$^{1}$,
 Jieru Zhao$^{2}$,
 Ming Liu$^{3}$,
 Zeke Wang$^{1}$\\

 \small $^{1}$ Zhejiang University\\
 \small $^{2}$ Shanghai Jiao Tong University\\
 \small $^{3}$ University of Wisconsin-Madison\\
}

\fancypagestyle{camerareadyfirstpage}{%
  \fancyhead{}
  
  \fancyhead[C]{
    \ifdefined\aeopen
    \parbox[][12mm][t]{13.5cm}{\hpcayear{} IEEE International Symposium on High-Performance Computer Architecture (HPCA)}    
    \else
      \ifdefined\aereviewed
      \parbox[][12mm][t]{13.5cm}{\hpcayear{} IEEE International Symposium on High-Performance Computer Architecture (HPCA)}
      \else
      \ifdefined\aereproduced
      \parbox[][12mm][t]{13.5cm}{\hpcayear{} IEEE International Symposium on High-Performance Computer Architecture (HPCA)}
      \else
      \parbox[][0mm][t]{13.5cm}{\hpcayear{} IEEE International Symposium on High-Performance Computer Architecture (HPCA)}
    \fi 
    \fi 
    \fi 
    \ifdefined\aeopen 
      \includegraphics[width=12mm,height=12mm]{ae-badges/open-research-objects.pdf}
    \fi 
    \ifdefined\aereviewed
      \includegraphics[width=12mm,height=12mm]{ae-badges/research-objects-reviewed.pdf}
    \fi 
    \ifdefined\aereproduced
      \includegraphics[width=12mm,height=12mm]{ae-badges/results-reproduced.pdf}
    \fi
  }
  \fancyfoot[C]{}
}
\begin{document}
\maketitle

\ifdefined\hpcacameraready 
  \thispagestyle{camerareadyfirstpage}
  \pagestyle{empty}
\else
  \thispagestyle{plain}
  \pagestyle{plain}
\fi

\newcommand{\hpcaheight}{0mm}
\ifdefined\eaopen
\renewcommand{\hpcaheight}{12mm}
\fi


\begin{abstract}

     The emerging microservice/serverless-based cloud programming paradigm and the rising networking speeds leave the RPC stack as the predominant data center tax. Domain-specific hardware acceleration holds the potential to disentangle the overhead and save host CPU cycles. However, state-of-the-art RPC accelerators integrate RPC logic into the CPU or use specialized low-latency interconnects, hardly adopted in commodity servers.

To this end, we design and implement \sysname, a software-hardware co-designed RPC on-NIC accelerator that enables reconfigurable RPC kernel offloading. \sysname{} connects to the server through the most widely used PCIe interconnect. 
To grapple with the ramifications of PCIe-induced challenges, \sysname{} introduces three techniques: (a) a target-aware deserializer that effectively batches cross-PCIe writes on the accelerator's SRAM using compacted hardware data structures; (b) a memory-affinity CPU-accelerator collaborative serializer, which trades additional host memory copies for slow cross PCIe-transfers; (c) an automatic field update technique that transparently codifies the schema based on dynamic reconfigure RPC kernels to minimize superfluous PCIe traversals. We prototype \sysname{} using the Xilinx U280 FPGA card. On HyperProtoBench, \sysname{} achieves an average of 2.3$\times$ lower RPC layer processing time than a comparable RPC accelerator baseline and demonstrates 2.6$\times$ achievable throughput improvement in the end-to-end cloud workload.

\end{abstract}
\section{Introduction}

\renewcommand{\thefootnote}{}
\footnotetext{$^{\dag}$Contributes equally}
\renewcommand{\thefootnote}{\arabic{footnote}}

Remote Procedure Call (RPC) is a paramount service block of today's cloud system stacks~\cite{erpc,dagger_asplos21,google_cloud_rpc_sosp23,grpc}. It abstracts remote computing resources and provides a simple and familiar programming model. Developers only prescribe type information for each remote procedure, and a compiler generates a stub code linked to an application to pass arguments via message. The RPC model has been widely adopted in many distributed applications, such as cloud storage~\cite{cloud_meets_rdma_nsdi21,smartds_isca23}, file systems~\cite{octopus+,kim2021linefs}, data analytics~\cite{zerochange_atc22}, consensus protocols~\cite{zhou2023electrode,zhou2021fault}, and machine learning systems~\cite{ray,pytorch_distributed}.


The RPC stack comprises two key components: (a) RPC protocol handling that parses the RPC headers, identifies the triggered message and the carried payload, and determines the target function; (b) serialization and deserialization, transforming between in-memory data fields and architecture/language-agnostic formats. A recent study from Google Cloud~\cite{google_cloud_rpc_sosp23} reports that the RPC processing occupies $\sim$7.1\% of CPU cycles across the entire fleet. Thus, it is important to accelerate the RPC execution, reduce this data center tax, and release more CPU cycles for revenue-generated applications.

Domain-specific hardware acceleration is a promising solution to build performant computing systems in the post-Moore's Law era. However, designing an RPC hardware accelerator is very challenging because the RPC stack is tightly coupled with the networking stack and application layer, whose processing should be efficiently streamlined into the data plane. As a result, researchers propose to use {\bf specialized on-chip interconnects} and closely integrate the RPC acceleration module in the host CPU chips~\cite{cereal_isca20, dagger_asplos21, optimus_asplos20, cerebros_micro21, protoacc_micro21}. For example, Cereal~\cite{cereal_isca20} introduces a special memory access interface to allow low-latency host memory accesses from the RPC accelerator. Dagger~\cite{dagger_asplos21} leverages Intel UPI~\cite{intel-upi} interconnect to facilitate RPC stack processing. 

Unfortunately, none of these proposals can be easily adopted on commodity servers due to the lack of interconnect support. RPC stack is continuously and rapidly evolving. For example, widely used gRPC~\cite{grpc} has 9 major releases over the last twelve months. As such, integrating RPC logic into the real host CPU lacks enough flexibility. Besides, developing a function- and performance-capable interconnect that can be integrated into a server system takes many years of engineering efforts, such as the ECI bus from the pioneering Enzian platform~\cite{cock2022enzian}. The emerging Compute Express Link~\cite{cxl-spec} looks promising, but its physical layer runs atop PCIe, yielding sub-microsecond access latency~\cite{cxl-micro23,li2023pond}, which cannot satisfy the latency requirement of the above accelerators. This leads to an interesting question: \textbf{How to accelerate RPC on top of de facto and predominant server interconnect, i.e., PCIe?}

In this paper, we design and implement \sysname{}, a software-hardware co-designed PCIe-attached accelerator for reconfigurable RPC offloading. \sysname{} colocates with the PCIe-attached NIC. \sysname{}' hardware part comprises three building blocks: (1) a target-aware deserializer that takes RPC requests, deserializes the messages, and forwards the results to the host or accelerator memory; (2) a memory-affinity serializer, which fetches computed data from both the host and accelerator memory, performs serialization, and fabricates the response; (3) programable computing units, dynamically offloading RPC computing kernels. 
In sum, \sysname{} is a low-profile immediately deployable PCIe-attached on-NIC accelerator with a software abstraction to load the RPC stack and related computing kernels on demand.


Building \sysname{} is non-trivial because of the high cross-PCIe overheads, jeopardizing the interaction performance between the RPC stack and other system layers. First, the RPC deserialization process needs to write deserialized results in a field-by-field scheme, whose throughput is bounded by the number of PCIe transactions. For example, our empirical evaluation using HyperProtoBench~\cite{protoacc_micro21} shows that this limitation can degrade the attainable deserialization throughput by 2.8$\times$ in geometric mean. \sysname{} proposes a target-aware deserializer that temporarily batches the deserialized fields within one RPC message in the accelerator's SRAM and performs cross-PCIe writes only when necessary. We realize this by designing two compacted hardware data structures (schema table and temp buffer) and revamping the deserialization process. 

Second, the RPC serialization process is hindered by the high PCIe latency. A nested RPC message or dereference field (strings/bytes/repeated/sub-messages) would require multiple memory accesses in a pointer-chasing manner since the memory location of the sub-fields can only be known after the parent's content is fetched. The sub-microsecond latency of PCIe would significantly increase the overall serialization time ($\sim$4.6$\times$ compared with an on-chip accelerator). \sysname{} designs a memory-affinity CPU-Accelerator collaborative serializer that trades additional host memory copies for slow cross-PCIe transfer. We introduce a lightweight pre-serialization phase to materialize the data layout on the host memory and facilitate the accelerator-side serialization execution. Besides, we leverage the memcpy (memory copy) engines~\cite{intel-dsa,dsa_benchmark} residing in modern CPUs~\cite{intel-spr} to alleviate host CPU usage for large fields' copy. 

Third, computation partition between host and RPC kernels within the RPC handler would cause suboptimal data placement and incur superfluous PCIe traversals. People eagerly co-locate domain-specific logic along with the RPC stack to maximize the hardware specialization benefits~\cite{cdpu_isca23,profiling_big_data_isca23,smartds_isca23,google_video_accel_asplos21,vbench_asplos18}. However, unlike on-chip cache-coherent interconnects, dynamic splitting computation logic across the host and accelerator over PCIe is inflexible and cause inferior data placement. Therefore, \sysname{} develops an automatic field update technique that transparently codifies the schema based on host/RPC kernel layout. As such, the accelerator deserializer can place the fields in suitable locations to avoid PCIe traversals. 

We built \sysname{} over an Xilinx Alevo U280 FPGA and evaluated it in several real-world scenarios. In a cloud image compression application, \sysname{} increases the achievable throughput by 2.6$\times$ and reduces the average (99th percentile) latency by 2.6$\times$ (1.9$\times$) compared with an RPC accelerator baseline. Using Google's HyperProtoBench~\cite{protoacc_micro21}, \sysname{} reduces the data serialization time by 4.3$\times$ in geometric mean. \sysname{} achieves similar performance as prior specialized on-chip accelerators from the literature. The source code will be open-sourced.

\section{Background and Motivation}
\label{sec:back-moti}


\subsection{Remote Procedure Calls}
\label{subsec:rpc}

A typical RPC layer consists of two execution logics: RPC protocol handling and de/serialization.

\squishlist
    \item {\it RPC protocol handling.}
    \cBlue{In the transmitting path (TX), the protocol handling mainly involves creating an RPC header. In the receiving path (RX), the protocol handling involves parsing the RPC header and dispatching the deserialized message to an idle CPU core to execute the target caller function.}
    This process is usually lightweight compared with RPC payload processing;
    
    \item {\it Serialization and deserialization.} Object serialization and deserialization are heavyweight operations and exist in RPC TX/RX, respectively. 
    RPC serialization transforms the in-memory fields into architecture and language-agnostic formats that can traverse through the network~\cite{zerializer_hotos21}. Deserialization operates conversely.
\squishend

\vspace{0.2\baselineskip}
\noindent
{\bf Protobuf.} We focus on the Protocol Buffer serialization library~\cite{protobuf}, widely used by many cloud applications.

\squishlist
    \item {\it Protobuf message definition.}
    It defines the logical transformation between the in-memory format and the wire format. A protobuf message is a collection of fields, usually called ``schema''. Each message field has a type, name, field number, and labels (e.g., ``repeated''). Field types can be (a) basic scalar types such as integers and strings; or (b) a nested user-defined message, also called a sub-message. Based on the memory layout, these fields can be further classified into two classes. One uses direct addressing, meaning that the value is within the memory location of its parent message, such as doubles and integers. The other uses indirect addressing (dereferences), indicating that their actual value is in a pointer-referenced memory location, such as strings, bytes, or sub-messages. In real-world applications the RPC message's depth can reach up to a dozen levels or more~\cite{protoacc_micro21};
    

    \item {\it Varint encoding of protobuf.}
    Data decoding/encoding is one of the most time-consuming operations in the de/serialization process~\cite{protoacc_micro21}, especially for small data fields. Data encoding widely exists in many popular serialization frameworks~\cite{protobuf, thrift} for message reduction.
    Protobuf uses variable-length integer encoding (known as varint). The encoding uses the most significant bit in each byte to indicate if the next byte is part of the same integer and the remaining 7 bits are used to store the actual value. Protobuf uses the tag-length-value format (TLV)~\cite{protobuf_tlv} for length-delimited fields (such as string and sub-message) and tag-value format (TV) for varints or fixed-length fields (such as double). 
    Handling these byte-wise and bit-wise operations on general-purpose modern CPUs is costly~\cite{zerializer_hotos21, cerebros_micro21, protoacc_micro21}, but can be easily accelerated via hardware specialization. 
    
\squishend

\subsection{Prior Hardware RPC Acceleration}

Researchers have developed several hardware-accelerated RPC solutions~\cite{cereal_isca20, dagger_asplos21, optimus_asplos20, cerebros_micro21, protoacc_micro21} to reduce the RPC stack processing overheads and save host CPU cycles. For example, Cereal~\cite{cereal_isca20} introduces a special memory access interface, enabling low-latency host memory access from the RPC accelerator. Dagger~\cite{dagger_asplos21} leverages a cache-coherent on-chip interconnect (UPI) to facilitate collaborative RPC stack processing between an FPGA-based SmartNIC and the host CPU. However, Dagger does not support nested structures or pointers, which are heavily used in today's applications. Optimus Prime~\cite{optimus_asplos20} and Cerebros~\cite{cerebros_micro21} place an on-chip accelerator for handling the de/serialization phase of the RPC stack. ProtoACC~\cite{protoacc_micro21} then develops a near-core de/serialization accelerator for Protobuf using a RISC-V SoC.

\captionsetup[table]{labelfont=bf, labelsep=period}

\begin{table*}
    \centering
    \scriptsize
    \begin{tabularx}{\linewidth}{>{\centering}m{2.5cm}||>{\centering}m{2.5cm}|>{\centering}m{1cm}|>{\centering}m{1.2cm}|>{\centering}m{5cm}|>{\centering\arraybackslash}m{3cm}}
     
        \hline
        \textbf{System} & \textbf{Interconnect} & \textbf{Latency} &  \textbf{Throughput} & \textbf{Accelerated RPC Stack} & \textbf{Accelerated RPC Kernels}\\
        \hline
        \hline
        
        \textbf{Cereal\cite{cereal_isca20}} &  MAI & 40 ns & 76.8 GB/s 
            & Customized De/Serialization & N/A \\ \hline
        
        \textbf{Optimus Prime\cite{optimus_asplos20}} & 2D mesh NoC & 45 ns & 64 GB/s 
            & Protobuf/Thrift-based De/Serialization & N/A \\ \hline
        
        \textbf{Cerebros\cite{cerebros_micro21}} & 2D mesh NoC & 45 ns & 64 GB/s 
            & Thrift-based De/Serialization, RPC protocol & N/A \\ \hline
        
        \textbf{ProtoACC\cite{protoacc_micro21}} & TileLink System Bus & 30 ns & N/A 
            & Protobuf-based De/Serialization & N/A \\ \hline
        
        \textbf{Dagger\cite{dagger_asplos21}} & Intel UPI & 125 ns & 19.2 GB/s 
            & Customized De/Serialization, RPC protocol, & Yes \\ \hline
            
        \textbf{\sysname{}} & {\bf PCIe} & {\bf 1250 ns} & {\bf 12.8 GB/s}
            & {\bf Protobuf-based De/Serialization, RPC protocol} & {\bf Yes} \\ \hline

    \end{tabularx}
    \caption{Hardware specification comparison.}
    \vspace{-4ex}
    \label{tab:prior_systems}
\end{table*}

\vspace{0.5\baselineskip}
\noindent
{\bf Limitation.} As shown in Table~\ref{tab:prior_systems}, all existing solutions require on-chip design or specialized interconnects for low latency. \cBlue{The main challenge of on-chip accelerator solutions is that integrating a specialized but not generalized function (e.g., RPC) into the commercial CPUs is very intrusive to the server CPU design. Considering that RPCs are evolving rapidly while the CPU design cycle takes years, integrating RPCs into the CPU chip is very costly and impractical.} 
As a result, none can be easily employed on commodity servers and immediately deployed at scale. 
{\bf Therefore, we aim to build an immediately deployable PCIe-attached RPC accelerator that offers the same programmability and comparable performance as prior on-chip designs.}

\vspace{-1ex}
\subsection{Challenges}
\label{subsec:challenges}


\begin{figure}
	\centering
	{\includegraphics[keepaspectratio=true, width=0.95\linewidth]{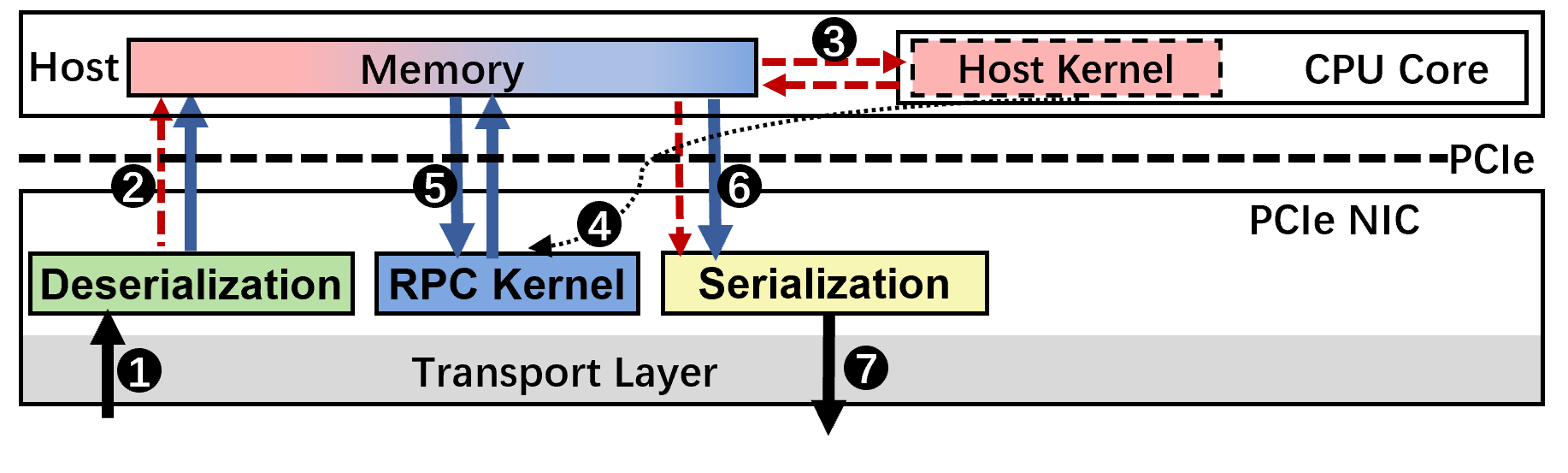} 
        \vspace{-2ex}
        \caption{High-level system architecture and request workflow of a PCIe-attached RPC accelerator.}
	\label{fig:dataflow}}
        \vspace{-4ex}
\end{figure}

\cBlue{Integrating a specialized function into the PCIe-attached NIC is much easier, takes much less engineering effort, and has been proven to be practical in many production systems. For example, Google integrates their transport protocol “Falcon” into the Intel IPU~\cite{google-falcon-ipu}, and AWS integrates storage function into their Nitro SmartNIC~\cite{aws_nitro}. 
} 

\cBlue{Anathor option is to implement a standalone PCIe-based RPC accelerator. However, both the transmitting and receiving paths incur multiple redundant cross-PCIe RPC message movements between the RPC accelerator and the NIC. Therefore, we prefer integrating the RPC accelerator in the NIC.}

Figure~\ref{fig:dataflow} sketches such a high-level design that offloads the RPC stack to the PCIe-attached NIC. When an RPC request arrives at the accelerator (\cycle{1}), the message is first deserialized via the hardware engine, where the deserialized fields are written into the host memory (\cycle{2}). Next, the host (\cycle{3}) and RPC kernel (\cycle{4} and \cycle{5}) are triggered collaboratively to process the RPC message. Then, the serialization engine retrieves the processing results, performs the serialization task (\cycle{6}), and sends back data through the network (\cycle{7}). The overall design seems straightforward but imposes three unique challenges.




{\bf C1: The limited number of concurrent PCIe transactions hinders the deserialization throughput.}
The deserialization engine writes back deserialized results into the host memory in a field-by-field manner~\cite{cerebros_micro21, protoacc_micro21}. However, these writing objects are small, incurring numerous DMA writes and small-sized PCIe transactions, quickly saturating the PCIe transaction rate. We built a deserialization accelerator (on the Xilinx U280 FPGA) based on ProtoACC~\cite{protoacc_micro21}, and enforced it to put deserialized results into either the host memory (crossing PCIe) or the FPGA off-chip memory. Our evaluations on HyperProtoBench~\cite{protoacc_micro21} show that cross-PCIe deserialization can only achieve 5.6$\times$ lower throughput compared with writing results to the FPGA's off-chip memory. 

{\bf C2: The high PCIe interconnect latency drastically decelerates the serialization performance.}
As described in $\S$\ref{subsec:rpc}, an RPC message can be a deeply nested memory object, causing multiple pointer-chasing memory accesses under retrieval during the serialization phase (\cycle{6}), which is extremely inefficient when crossing PCIe (taking sub-microseconds). 


\begin{figure}
	\centering
	{\includegraphics[keepaspectratio=true, width=0.9\linewidth]{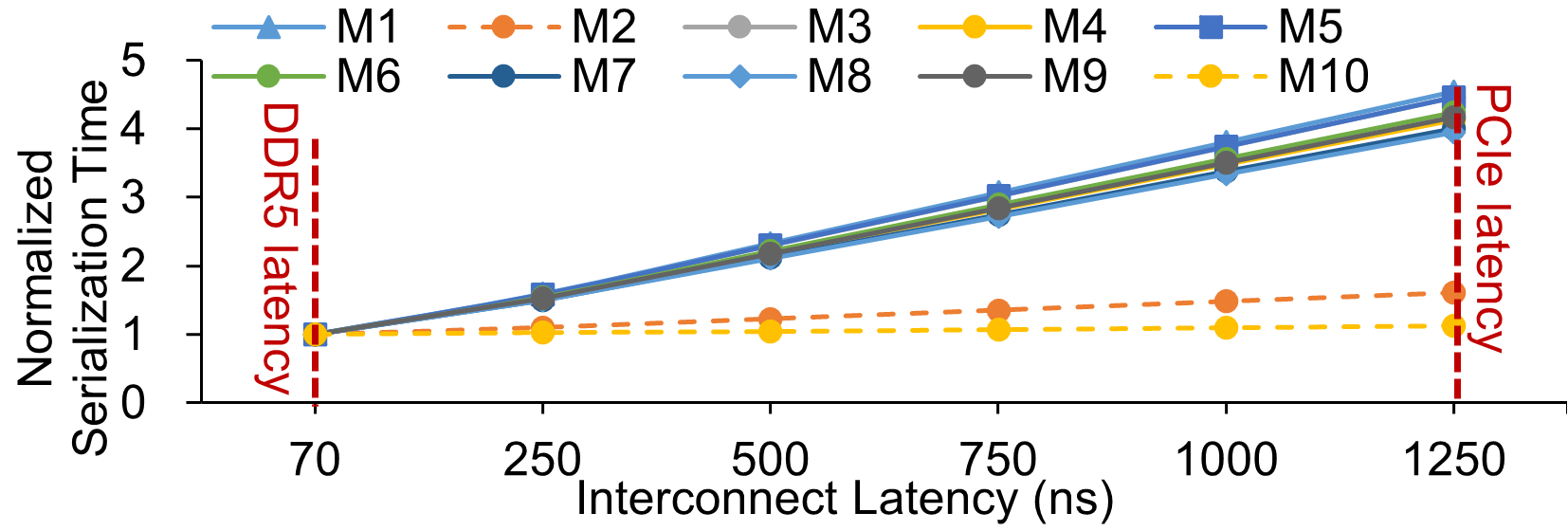} 
        \caption{Normalized serialization time when increasing the simulated PCIe latency for different messages.}
	\label{fig:e_motivation_seriail_latency}}
        \vspace{-4ex}
\end{figure}

\label{evaluation:serialization_loss}
To illustrate this, we implement a protobuf serialization accelerator based on ProtoACC~\cite{protoacc_micro21}. We measure the serialization time when varying interconnect latency through the Xilinx Vivado simulator~\cite{vivado}. Figure~\ref{fig:e_motivation_seriail_latency} illustrates the normalized serialization time for all messages of Bench2 in HyperProtoBench~\cite{protoacc_micro21}. As expected, when the interconnect latency increases from DDR5's 70ns latency to commercial PCIe's 1250ns, the end-to-end serialization's time increases by 3.4$\times$ in geometric mean, due to the complex nested message structure. 
The exception is that the two messages (M4 and M10) only present a marginal increase. This is because when the RPC message becomes large and flat (1.6MB and 0.6MB), the serialization performance is dominated by the data transfer time and is not sensitive to the interconnect latency.

{\bf C3: Suboptimal data placement causes superfluous PCIe accesses from host/RPC kernels.}
People eagerly offload domain-specific logic along with the RPC stack for core savings and performance maximization~\cite{dagger_asplos21,cdpu_isca23,profiling_big_data_isca23,smartds_isca23,google_video_accel_asplos21,vbench_asplos18}. These offloadable kernels are generally parallel-friendly with less data dependency. For example, researchers place a data compression engine for the cloud block storage application~\cite{smartds_isca23}. Ideally, one should divide incoming data and place them accordingly, such that the host and the offloaded RPC kernels only access their data locally. However, in reality, since the offloaded kernels (\cycle{5}) are only part of the RPC handler and the offloaded kernels may dynamically change in a multi-tenant environment, it becomes extremely challenging to design a clean and optimal partition. A suboptimal data placement is very costly for a PCIe accelerator considering its high latency for cross-PCIe traversals.


To illustrate this, we develop an RPC-based network function accelerator that co-locates with a PCIe-attached NIC. It serves as the cloud gateway~\cite{luoshen_nsdi24}. It performs L2/L3 protocol processing, network address translation (NAT), and packet de/encryption. We explore different computing-driven data placement strategies and find out that the worst-case placement can decrease the achievable throughput by 2.2$\times$ than the best-case placement.

\vspace{-1ex}
\section{\sysname: Design and Implementation}





\subsection{Overview}

\begin{figure*}
    \centering
    {\includegraphics[keepaspectratio=true, width=0.9\linewidth]{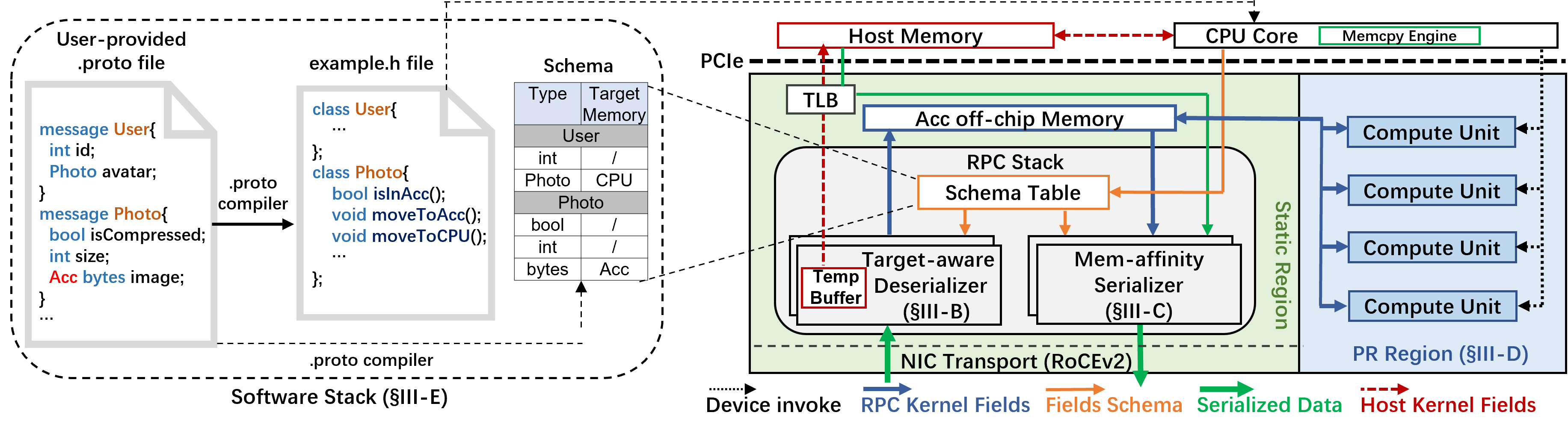}} 
    \caption{Software stack and hardware architecture of \sysname{}.}
    \vspace{-3ex}
    \label{fig:arch}
\end{figure*}

\sysname{} is a software-hardware co-designed PCIe-attached accelerator that allows offloading user-defined RPC kernels. Figure~\ref{fig:arch} provides the system overview. The hardware part consists of 1) a target-aware deserializer ($\S$\ref{subsection_deserialization}) that takes RPC requests, deserializes the messages, and forwards the results to the host or accelerator; 2) a memory-affinity serializer ($\S$\ref{subsection_serialization}), which fetches computed data, performs serialization, and fabricates the response; 3) programable computing units ($\S$\ref{cu}), dynamically offloading RPC computing kernels; and 4) a transport layer. \cBlue{RPCAcc adopts a RoCE-based transport layer~\cite{strom_eurosys20}, which is entirely offloaded to the NIC just like an RDMA NIC. The RPC acceleration logic is in the NIC and sits between the transport layer and the PCIe controller. When the RPC message is fabricated in the NIC RPC layer, the hardware will send the message using an “RDMA Send” verb and the remote side uses an “RDMA Recv” verb to receive incoming RPC requests. The benefit of putting the RPC acceleration and transport layer together in NIC hardware is that it avoids redundant data movement or PCIe traversals between the transport processing and RPC processing. }

We place (1), (2), and (4) in the static region of the board, while (3) in the partial reconfiguration region. Our software stack ($\S$\ref{sec:software} and $\S$\ref{subsection_field}) of \sysname{} consists of (a) a compiler that takes the user-defined RPC message specification and outputs both the message structure and accelerator-friendly configurations; and (b) a rich set of APIs to describe the RPC kernel task.

\vspace{-1ex}
\subsection{Target-aware Deserializer}
\label{subsection_deserialization}
To address challenge \#1 ($\S$\ref{subsec:challenges}), we develop a target-aware deserialization engine that forwards deserialized fields to the host or accelerator memory accordingly. Our deserialization logic has 4 independent computing lanes (i.e., deserializers). Each deserializer processes RPC requests one by one. Each deserializer executes the deserialization logic and converts the results into in-memory C++ objects.

We introduce two hardware data structures (described below) and revamp the deserialization process based on them. 

\squishlist
    \item {\bf Schema Table.} It is an SRAM region that stores the message structure of incoming RPC messages. For each field of an RPC class, we use one bit to indicate its target location type for the deserialized results. Fields used by the offloaded RPC kernel (host kernel) are forwarded to the accelerator off-chip memory (host CPU memory). The ``Schema Table'' is shared by all deserializers and serializers. Section~\ref{sec:software} describes how this bit is set and how the ``Schema Table'' is constructed;

    \item {\bf Temp Buffer.} Deserialized results used by RPC kernels are directly written to the accelerator off-chip memory. For others, we use a per-deserializer SRAM buffer (``Temp Buffer'') to store the deserialized fields temporarily. The buffer is 4KB and operates in an append-only mode, simplifying the buffer management. 
    The buffer size is configurable.
    When the buffer is full or the current RPC request's deserialization is finished, the deserializer triggers a DMA write and copies data to the intended host CPU memory. We call this batching mechanism ``One-shot DMA write''. Note that this batching mechanism would barely increase the deserialization latency, 
    since it only batches the fields within an RPC request instead of batching fields from different requests.
    
    
    

\squishend

\vspace{0.2\baselineskip}
\noindent{\bf Deserialization Procedure.} An incoming RPC message is assigned to an idle deserializer or buffered when there are no idle deserializers.
To avoid software allocation overhead, we reserve a host CPU memory region and an accelerator off-chip memory region for the deserializer. These two memory regions are divided into 4KB \footnote{\cBlue{4KB chunks lead to a small allocation time (0.68\% of the total deserialization time) and a small fragmentation overhead (3.6\% in HyperProtoBench). The size is configurable at system initialization and the users can choose a suitable value that balances both allocation time and memory fragmentation.}} chunks. 
There is a \cBlue{16K-entry TLB}\footnote{\cBlue{ We adopt a simple TLB implementation, which can only store pages with contiguous virtual addresses. 16K entries only occupy 0.29\% of the total SRAM resources in our FPGA prototype.}} to perform host CPU memory address translation on the accelerator. We use two SRAM-based FIFOs (called free-list FIFOs) to store the free chunks of the host/accelerator memory region. Allocating/freeing memory is translated to poping/pushing a 4KB chunk from/into the FIFO, simplifying hardware complexity. 

The deserializer first pre-allocates a host CPU memory chunk and an accelerator off-chip memory chunk. It then parses the RPC header to obtain the RPC message class ID and message length and queries the ``Schema Table'' based on the class ID, which returns the schema of this message class.
The accelerator then deserializes the message data in a field-by-field manner accordingly. When encountering a dereference sub-message,  the deserializer pushes the current message schema into an SRAM-based stack and deserializes the sub-message recursively.


During the deserialization, each deserialized field has one of the two target locations:
\squishlist
    \item \textbf{Host CPU memory:} The deserialized result is assigned a CPU memory location from the pre-allocated CPU memory chunk. As described above, we would temporarily save it in the ``Temp Buffer''.

    \item \textbf{Accelerator off-chip memory:} 
    The deserialized data is assigned an accelerator memory location from the pre-allocated off-chip memory chunk. Then the result would be directly written to this location. The corresponding field pointer in the parent message would be updated to point to this off-chip memory location. 
\squishend

When the deserializer exhausts the pre-allocated chunks, it allocates a new chunk from one of the two free-list FIFOs. When exhausting pre-allocated host CPU memory chunks, the deserializer additionally uses a DMA write to flash the ``Temp Buffer'' into the corresponding host CPU memory.
Upon deserialization completion, the accelerator then notifies the host CPU of an incoming RPC message. 


\textbf{Summary.} Compared with the traditional field-by-field deserialization scheme~\cite{protoacc_micro21, cerebros_micro21, optimus_asplos20}, our target-aware deserializer uses effective batching and reduces unnecessary PCIe traffic by storing certain fields in the accelerator local off-chip memory. Besides, we directly store the fields that are not needed by the host CPU in the NIC accelerator's off-chip memory, greatly reducing unnecessary PCIe transactions. 


\vspace{-1ex}
\subsection{Memory-affinity Serializer}
\label{subsection_serialization}
We next discuss how to address challenge \#2 ($\S$\ref{subsec:challenges}) in \sysname{}. There are two general serialization design choices:

\begin{figure}
	\centering
	{\includegraphics[keepaspectratio=true, width=0.99\linewidth]{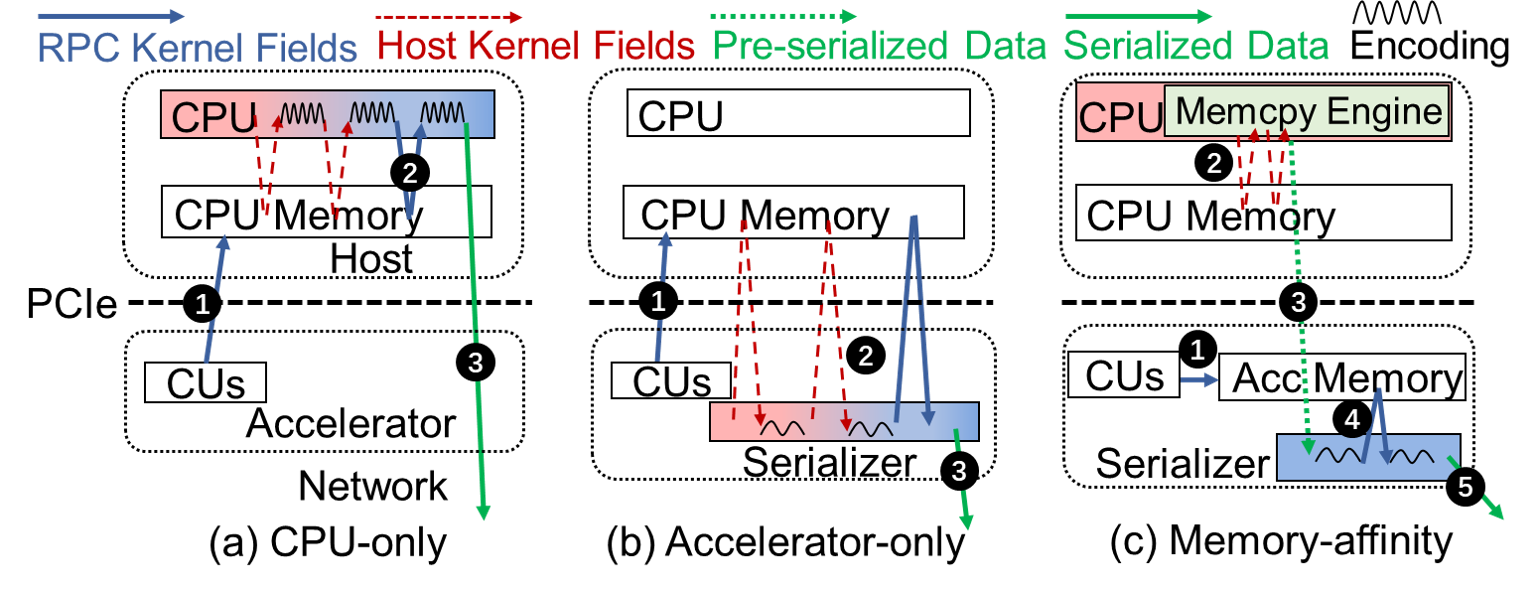}} 
        \caption{Compaison of three serialization strategies.}
	\label{fig:serialization_dataflow}
        \vspace{-3ex}
\end{figure}

\squishlist
    \item {\bf Option\#1: CPU-only Serialization.} Figure~\ref{fig:serialization_dataflow}-a depicts the process of  CPU-only serialization. Upon the compute unit (CU) finishing computation, it writes the result back to the host CPU memory (\cycle{1}). The CPU then retrieves all the fields and serializes them (\cycle{2}), writing them into a DMA-safe memory region. At last, the accelerator reads data from the DMA-safe region, fabricates the RPC response, and sends it to the network (\cycle{3}). As CPU memory access latency is very low ($\sim$70ns), this approach can tolerate nested RPC messages well. However, it wastes host CPU cycles on CPU-inefficient encoding, while wasting PCIe bandwidth (GB/s, not transaction rate) drastically in stage \cycle{1};

    \item {\bf Option \#2: Accelerator-only Serialization.} Figure~\ref{fig:serialization_dataflow}-b shows how an on-NIC accelerator performs serialization independently. Compared with CPU-only serialization, the difference is that the serialization is fully offloaded to the NIC hardware.
    Accelerator-only serialization consumes minimal host CPU cycles. 
    However, as discussed in $\S$\ref{subsec:challenges}, the high cross-PCIe latency would jeopardize the serialization time, especially for deeply nested RPC messages. 

\squishend

{\bf Our approach: Memory-affinity CPU-Accelerator Collaborative Serialization (Option \#3).} To address the limitations of the above two, we distribute the serialization logic across the host CPU and accelerator, aiming to achieve the best of two worlds:
minimizing PCIe transfers while consuming the fewest host CPU cycles. 
Our key idea is to add a lightweight CPU pre-serialization phase to the host to materialize the data layout for fields residing in the host memory, which trades additional fast host memory copies for slow PCIe accesses.
Figure~\ref{fig:serialization_dataflow}-c highlights the process.
The compute unit writes results into the accelerator memory instead of the host memory (\cycle{1}). Then the host CPU retrieves all local fields and pre-serializes them (\cycle{2}) without CPU-inefficient encoding. Next, the CPU sends the pre-serialized data to the serializer (\cycle{3}), which encodes these data and further serializes fields that reside in the accelerator memory (\cycle{4}). At last, the serialization module merges the CPU and accelerator memory fields and sends the merged result out to the network as an RPC response (\cycle{5}). 
\cBlue{Modern server CPUs~\cite{intel-spr} are integrated with on-chip memcpy engines (Data Stream Accelerator~\cite{intel-dsa, dsa_benchmark}) and the pre-serialization process offload the copies of large fields to the memcpy engines to save host CPU cycles during the pre-serialization.}.  


Next, we discuss the detailed serialization procedure:

\vspace{0.2\baselineskip}
\noindent{\bf Stage 1: CPU Pre-serialization.} We maintain a small DMA-safe buffer to store the CPU pre-serialization output. The process iterates the to-be-serialized object in a field-by-field manner.
The process scans each encountered field and writes the non-contiguous results into the contiguous DMA-safe buffer. Upon finishing, the software uses an MMIO write to notify the accelerator of the address and length of the pre-serialized data, and other required information to construct an RPC header.
The pre-serialization has three unique properties that can help reduce CPU cycles:

\squishlist
    \item {\bf Memcpy Offload. } 
    The copy of large CPU memory fields in the host CPU memory can be offloaded to the memcpy engines. The CPU asynchronously invokes the memcpy engines to reduce the required CPU cycles at most. 

    \item {\bf Encoding Offload. } The pre-serialization process would not perform CPU-inefficient encoding, deferred to the hardware accelerator.

    \item {\bf Skipping Accelerator Fields. } 
    The pre-serialization process only pre-serializes fields residing in the host CPU memory. If encountering a field residing in accelerator memory, it only writes the pointer value and data length into the DMA-safe buffer. 

\squishend

 
\vspace{0.2\baselineskip}
\noindent{\bf Stage 2: Accelerator Serialization.} When the accelerator is notified after the completion of CPU serialization, the accelerator constructs an RPC header in an SRAM region, which is called ``TX Arena'' and is used to store the final RPC message that is to be sent to the network. The hardware serializer then uses a DMA read to fetch the pre-serialized data from the host CPU memory, iterates pre-serialized data, and performs varint encoding. The serializer encodes the pre-serialized data in a per-512-bit manner. For each 512-bit, the encoding can be done within one cycle. 
If a pointer referring to the accelerator off-chip memory is found, the serializer reads the referred data from the accelerator off-chip memory, serializes it, and writes the result into the ``TX Arena''. 
After fetching the data from the accelerator off-chip memory, the engine serializes it, writes the result to the corresponding address in the ``TX Arena'', and continues the iteration. When the iteration finishes, the RPC header, and the serialization results now lie contiguously in the ``TX Arena''. The transport layer can transmit these data into the network.

\textbf{Summary.} Our memory-affinity CPU-accelerator collaborative serializer trades fast host memory copies for slow cross-PCIe transfer. It introduces a lightweight pre-serialization phase to materialize the data layout and facilitate the accelerator-side serialization execution. To further reduce CPU overhead during pre-serialization, we leverage memcpy engines to perform data copies of large fields. 


\begin{table*}[t]
    \centering
    \scriptsize
    \begin{tabularx}{\textwidth}{>{\centering}m{2cm}|m{4.5cm}| m{10cm}}
       
         \hline

        \textbf{Member Function} & \textbf{Parameter} & \textbf{Description}\\
         \hline
         \hline

         \small{.program()}& The file location of the bitstream (\texttt{bitFilePath}), 
                            the programmed RPC kernel (\texttt{kernelType}) &
         Program the compute unit with the provided bit file; the compute unit is labeled as ``kernelType''.
         \\
        \hline

         \small{.getType()} & N/A &
         Return the string of the labeled ``kernelType''.
         \\
        \hline

         \small{.submitTask()}& Address of input to the CU (\texttt{inputAddr}), input size in bytes (\texttt{inputSize}), address of output result (\texttt{outputAddr}), output size in bytes (\texttt{outputBufSize})&
         Submit a new task to the compute unit, and an asynchronous task event will be returned.  The CU would fetch inputSize bytes from the accelerator memory address inputAddr. When the engine completes, it writes the result into the accelerator memory address outputAddr.
         \\
        \hline

         \small{.poll()}& The completion signal of the compute unit (\texttt{taskEvent}) &
         Busy polling the taskEvent until it completes.
         \\

         \hline
    \end{tabularx}
    \caption{Member functions of compute units in \sysname{}.}
    \vspace{-2ex}
    \label{tab:api}
    
\end{table*}

\vspace{-1ex}
\subsection{Compute Unit}
\label{cu}
In addition to accelerating the RPC stack itself, \sysname{} allows user to program the compute units (CUs) with their hardware logic to further offload compute-intensive computations in the RPC requests. We call these offloaded computations as RPC kernels. A compute unit in \sysname{} is a partially reconfigurable FPGA block. Each CU has a memory interface connected to the accelerator off-chip memory.

CUs interact with the host software uniformly and provide a set of APIs (Table~\ref{tab:api}).
Each CU has a descriptor ring in the accelerator SRAM and a notification ring in the host CPU memory. To activate a CU, the host software has to submit a descriptor to the descriptor ring using an MMIO write (``submitTask()''). After submission, the address of an available entry in the notification ring would be returned to the software. Submitting computation tasks is an asynchronous process, and the software can poll the returned address to be aware of this task's completion (``poll()''), akin to the BlueFlame~\cite{blueflame} mechanism in the Mellanox NICs.

\squishlist
    \item {\bf Descriptor ring.} Entries in the descriptor ring are submitted by the host CPU software, where each entry consists of the input address, input length, output address, and output buffer size. When a CU becomes idle, it fetches the next ready descriptor from the ring, and reads data from the off-chip memory using the input address and input length;

    \item {\bf Notification ring.} When computation finishes, the CU first writes the results into the output address. Then the result length and the completion signal would be written into the corresponding notification entry in the host CPU memory using one DMA write.
\squishend
In addition to submitting tasks to the CU and polling the completion, the user can use ``.program()'' to reprogram the CU with a given FPGA bit file and use ``getType()'' to check the currently supported computation of the compute unit. \cBlue{In the current implementation, we create four PR blocks of the same size. Equal-size PR blocks expose limitations on flexibility. It's also possible to leverage the techniques proposed in prior works~\cite{vital_asplos20, heterovital_isca21, amorphos_osdi18} to dynamically manage the PR region and this could be our future work.}

\vspace{-1ex}
\subsection{\sysname{} Software Stack}
\label{sec:software}

\sysname{} provides a compiler toolchain and programming APIs that enable dynamic and reconfigurable RPC stack and computing kernel offloading.


\subsubsection{\sysname{} Compiler}

Our compiler takes a user-provided \texttt{.proto} file that contains the RPC message structure, and generates (1) a header file for applications running on the host CPU and (2) a schema definition stored in the ``Schema Table'' of the \sysname{} accelerator for orchestrating RPC request flow.

\sysname{} fully supports Protobuf3~\cite{protobuf} format. Programmers first define the RPC message format in the \texttt{.proto} file and specify a field with labels such as ``optional'' and ``repeated''. In \sysname{}, the user can additionally specify a dereference field (string/bytes/repeated/sub-message) with a custom label ``Acc''\footnote{``Acc'' represents the accelerator's off-chip memory.}. 
The compiler first scans the \texttt{.proto} file, recording the structures of each message and attribute of each field. Then the compiler generates a header file and a schema definition based on the scanned results. 

The header file mainly consists of generated RPC message classes including de/serialization functions and three unique member functions (i.e., ``moveToAcc'', ``moveToCPU'', and ``isInAcc'') for each dereference field (Table~\ref{tab:mem_function}).
``isInAcc'' checks whether the pointer refers to the accelerator off-chip memory.
``moveToAcc'' moves data from the pointed CPU memory to the accelerator off-chip memory while ``moveToCPU'' operates reversely. 

The schema definition stores the RPC message structure and the attributes of each field. The schema definition is stored in the accelerator ``Schema Table''. During the deserialization, the deserializer selectively puts the deserialized field in the host/accelerator memory based on whether the field has an ``Acc'' label.

\begin{table}[t]
    \centering
    \scriptsize
    \begin{tabularx}{\linewidth}{>{\centering}m{2cm}||m{5.5cm}}
     
        \hline
        \textbf{Member Function} & \textbf{Description}\\
        \hline
        \hline
         
         \small{.isInAcc()}& 
         Check whether the data is in the accelerator memory.
         \\
         \hline

         \small{.moveToAcc()}& 
         Move data to the accelerator memory, and the field's pointer will be updated to this new location.
         \\
         \hline

         \small{.moveToCPU()}&
         Move data to the host CPU memory, and the field's pointer will be updated to this new location.
         \\

         \hline
    
    \end{tabularx}
    \caption{Dereference fields functions in \sysname{}.}
    \vspace{-3ex}
    \label{tab:mem_function}
\end{table}

\subsubsection{\sysname{} Programming Interface}


We take an RPC-based compression accelerator as an example. Listing~\ref{listing:example} shows the pseudo implementation. 
The RPC request message (User) and response message (Photo) are defined in Figure~\ref{fig:arch}. 
This application is representative as it involves a lightweight host kernel (authorization) and a compute-intensive RPC kernel (compression). The authorization only involves lightweight processing and usually has many data dependencies. The compression is compute-intensive, easy to parallelize, and does not have data dependencies. 

Next, we show how to realize the application using the provided programming interfaces assuming that the compression bit file is ready and has been programmed in the CU. When an RPC request arrives, the software first checks whether the user is authorized (L1). After authorization, the software checks whether the CU can perform compression (L4).

\squishlist
    \item If so (L5-10), the application first ensures that the raw avatar data of the request is put in the accelerator memory (L5). Otherwise, it moves the data to accelerator memory (L6). It then invokes the compute unit to execute the compression RPC kernel (L8), poll the result (L9), and set the size in the RPC response (L10);

    \item Otherwise (L13-16), we first ensure that the raw avatar data is in the CPU memory (L13). If not, we move the data to the host memory (L14). We then perform CPU compression and set the compressed image size (L16).
\squishend
Finally, the RPC response is fabricated (L18).
\sysname{} allows developers to focus on high-level application logic instead of dealing with the RPC layer and host-accelerator interactions.



\begin{algorithm} []
	\SetAlFnt{\tiny} \linespread{1.0} \selectfont \caption{\sc \sysname{} programming example} 
	\label{listing:example}
	\SetKwInOut{Define}{Define}
	\newcommand\mycommfont[1]{\ttfamily\textcolor{blue}{#1}}
    \SetCommentSty{mycommfont}
	
	\begin{footnotesize}
	    
		\Define{
		    $req$: RPC request,\\
                $res$: RPC response,\\
                $cu$: compute unit}                
		\If{isIdAuthorized($req.id$) == false} {
            \tcc{Host kernel, omitted}
        }
        reqData = req.avatar.image;\\
        \If{cu.getType() == "compress"} {
            \If{reqData.isInAcc() == false}{
                reqData.moveToAcc();
            }
            e = cu.submitTask(...);\\
            poll(e);\\
            res.size = e.size;\\
        }\Else{
            \If{reqData.isInAcc() == true}{
                reqData.moveToCPU();
            }
            res.size = compressOnCPU(...);\\
        
        }
        return $res$;
     \vspace{-1ex}
	\end{footnotesize}
\end{algorithm}

\vspace{-2ex}
\subsection{Automatic Field Updating}
\label{subsection_field}


At default, developers manually assign ``Acc'' labels for data fields, indicating that they are likely to be used by a compute unit in the accelerator. This approach works well for pre-known request traffic because one can profile the data access pattern and devise an optional data placement scheme. However, such an assumption is no longer held in the cloud setting~\cite{solar_sigcomm22, smartds_isca23}, where computing demand varies continuously. Thus, developers would reprogram the computing unit and determine which RPC kernels would benefit the most from hardware acceleration, completely breaking the established data partition layout and causing unnecessary PCIe traversals across two computing domains.


We instead propose an automatic field updating mechanism that allows modifying message schema at runtime. Specifically, it automatically adds/removes the ``Acc'' label for all dereference fields at runtime when calling ``moveToAcc'' and ``moveToCPU'' member functions. 

When ``moveToAcc'' or ``moveToCPU'' is invoked, not only an MMIO would be issued to notify the accelerator to move the field's content across PCIe, but also the corresponding entry in the ``Schema Table'' would be updated. If ``moveToNAcc'' is called, it indicates that the field should be added with a ``Acc'' label, while ``moveToCPU'' would remove the ``Acc'' label of this field. In this way, when the next same RPC arrives, the deserializer can use the updated ``Schema Table'' to deserialize RPC messages, thus avoiding redundant data movement within subsequent RPC requests. 

\cBlue{ {\bf Limitation.} If a field is needed by both the CPU and the \sysname{} compute unit, the field must be fetched over the PCIe bus. This is unavoidable in the current design as the CPU and PCIe device are not in a coherent domain. However, we believe this is relatively rare as users should try to ensure that the offloaded function is a standalone logic and does not reuse the CPU-side data.}

{\bf Summary.} Automatic field updating poses three benefits. First, it frees users from the tedious task of manual and explicit assignment of ``Acc'' labels in .proto file. Second, our .proto file can be completely the same as Protobuf3. As such, existing applications can easily adopt \sysname{} without modification of .proto files. Third, it can accommodate partial reconfiguration's dynamic offloading. When the RPC kernels change from CPU execution to compute unit execution or reversely, the deserializers can place the corresponding fields correctly after one incorrect placement. 

\vspace{-1ex}
\section{Evaluation}

Our evaluations aim to answer the following questions:
\squishlist
    \item How effective is the target-aware deserializer ($\S$\ref{evaluation:deserialization})?
    \item How effective is the memory-affinity serializer ($\S$\ref{evaluation:serialization})?
    \item How does the performance of \sysname{} compare to that of an SoC SmartNIC ($\S$\ref{subsec:eval-soc}), Dagger ($\S$\ref{subsec:eval-dagger}), and an on-chip accelerator ($\S$\ref{evaluation:comparison_with_on_chip_acc})?
    \item How can automatic field updating address the data layout problem under compute reconfiguration ($\S$\ref{subsec:eval-auto})?
    \item How much performance acceleration can \sysname{} achieve for RPC-based workloads ($\S$\ref{subsec:eval-end})? 
    \item How many resources do the accelerator use ($\S$\ref{subsec:eval-hw})?
\squishend

\vspace{-1ex}
\subsection{Experimental Setup}
\noindent{\bf Hardware Testbed.} Our hardware testbed consists of two servers, each having two 16-core Xeon Silver 4514Y CPUs running at 2.0GHz, 512 GiB (8x64 GiB) 4800 MHz DDR5 memory, and a 60 MiB LLC. Each core has a memcpy engine (Intel DSA~\cite{intel-dsa, dsa_benchmark}). Each server is equipped with a Mellanox Dual-port ConnectX-5 100 Gb NIC ($\times$16) and a Xilinx U280 ($\times$16)~\cite{u280} FPGA which features an 8 GiB off-chip high bandwidth memory (HBM).

\vspace{0.2\baselineskip}
\noindent{\bf CPU Baseline. } We implement the baseline ``CPU-only'' (2K LoCs of C++) that runs the entire RPC stack and all computations on the CPU. It uses the DPDK-based eRPC~\cite{erpc} library and adopts Protobuf3~\cite{protobuf} as the de/serialization format. \cBlue{It is well-optimized under optimization methods such as zero-copy, huge page, kernel-bypass, and polling mode driver.}

\vspace{0.2\baselineskip}
\noindent{\bf ProtoACC-OnChip Baseline.} As no on-chip RPC accelerator hardware exists, we implement the ``ProtoACC-OnChip'' simulation baseline based on ProtoACC~\cite{protoacc_micro21}. We mainly modify the frequency to 250 MHz /2 GHz and set the memory access latency to 70 nanoseconds (same level as our host CPU memory latency).


\vspace{0.2\baselineskip}
\noindent{\bf ProtocACC-PCIe Baseline. }We implement baseline ``ProtocACC-PCIe'' (3K LoCs of C++ and 3.7K LoCs of Chisel3) and prototype it on U280 FPGA hardware. It offloads the entire RPC stack and computation kernels to the hardware. The FPGA performs the de/serialization logic following ProtoAcc~\cite{protoacc_micro21}, the difference is that our implementation uses PCIe interconnect. The serialization presedure is similar to the ``Accelerator-only'' approach as introduced in \S~\ref{subsection_serialization}. The RPC protocol is close to eRPC~\cite{erpc} and the transport layer adopts a modified version of Strom~\cite{strom_eurosys20}. The RPC/transport logic runs at 250 MHz.

\vspace{0.2\baselineskip}
\noindent{\bf BF3 Baseline.} We implement baseline ``BF3'' that offloads the RPC layer to SoC SmartNIC Bluefield-3~\cite{bf3}. The software stack is the same as the CPU baseline.

\vspace{0.2\baselineskip}
\noindent{\bf \sysname{} Implementation.} We prototype \sysname{} in the U280 FPGA (3K LoCs of C++ and 3.5K LoCs of Chisel3). The RPC protocol is similar to eRPC~\cite{erpc}. The transport layer adopts a modified version of StRoM~\cite{strom_eurosys20}. It features target-aware deserialization and memory-affinity CPU-Accelerator co-serialization. The RPC/transport logic runs at 250 MHz. 

\vspace{-1ex}
\subsection{Target-aware Deserializer}
\label{evaluation:deserialization}
\begin{figure}[]
    \subfloat[All messages]{
        \label{fig:oneshot_dma}
        \includegraphics[width=0.48\linewidth]{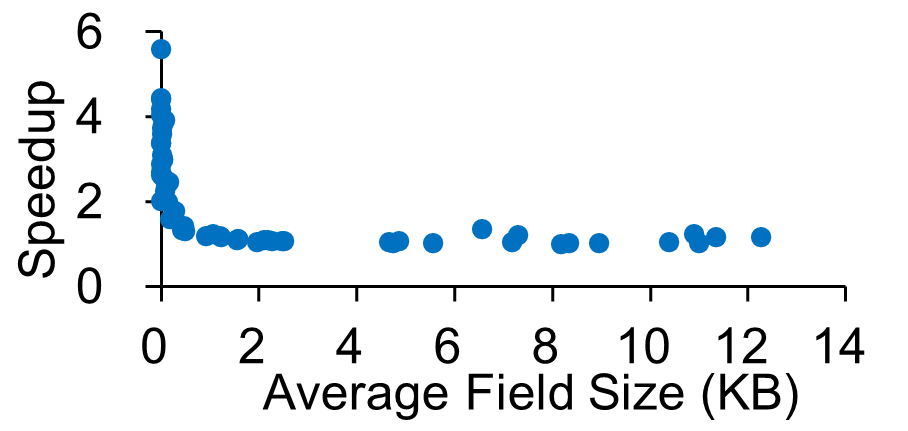}
    }
    \subfloat[Messages with small fields ($\leq$1KB)]{
        \label{fig:oneshot_dma_0_1KB}
        \includegraphics[width=0.48\linewidth]{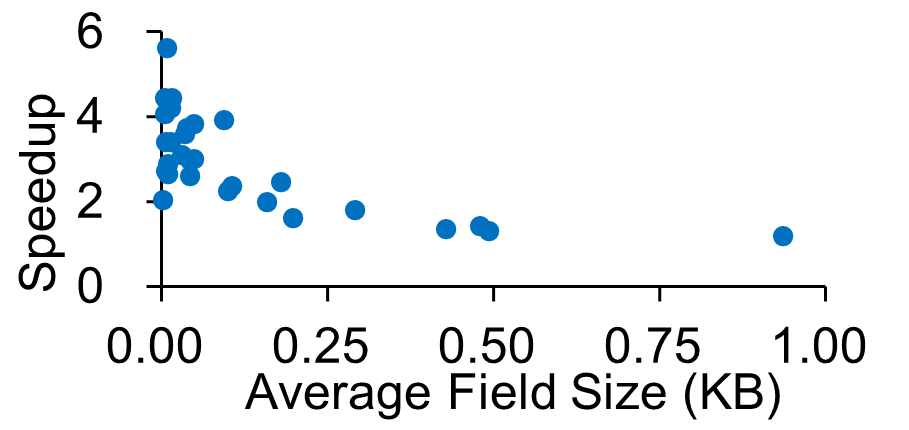}
    } 
    \vspace{-1ex}
    \caption{Throughput speedup of one-shot DMA write.} 
    \vspace{-3ex}
    \label{fig:e_oneshot_dma} 
\end{figure} 

    

We first examine the performance of our target-aware deserializer. We set all RPC data fields' destinations to the host memory and compare \sysname{} with the conventional field-by-field one. Figure~\ref{fig:e_oneshot_dma} reports the deserialization throughput improvement of ``one-shot DMA write'' over ``field-by-field'' when running HyperProtoBench~\cite{protoacc_micro21} (including six workloads, each containing 10 messages) for messages with different average field sizes. On average, \sysname{} outperforms the field-by-field solution by 2.2$\times$. 
For messages with small-sized fields (less than 1KB), \sysname{} achieves a higher speedup (3.1$\times$) because the field-by-field solution suffers from inefficient DMA that transfers numerous small objects. Instead, our one-shot DMA write scheme can combine small DMA writes into one large contiguous DMA write, yielding higher PCIe link utilization. 


\vspace{-1ex}
\subsection{Memory-affinity Serializer}
\label{evaluation:serialization}

We validate the effectiveness of the memory-affinity CPU-Accelerator collaborative serialization scheme by measuring the CPU cycle savings and serialization time. We use HyperProtoBench~\cite{protoacc_micro21} and five representative microservices in DeathStarBench~\cite{deathstarbench_asplos19} as workloads.

\begin{figure}[]
    \subfloat[HyperProtoBench]{
        \label{fig:serialization_offload_hyperproto}
        \includegraphics[width=0.48\linewidth]{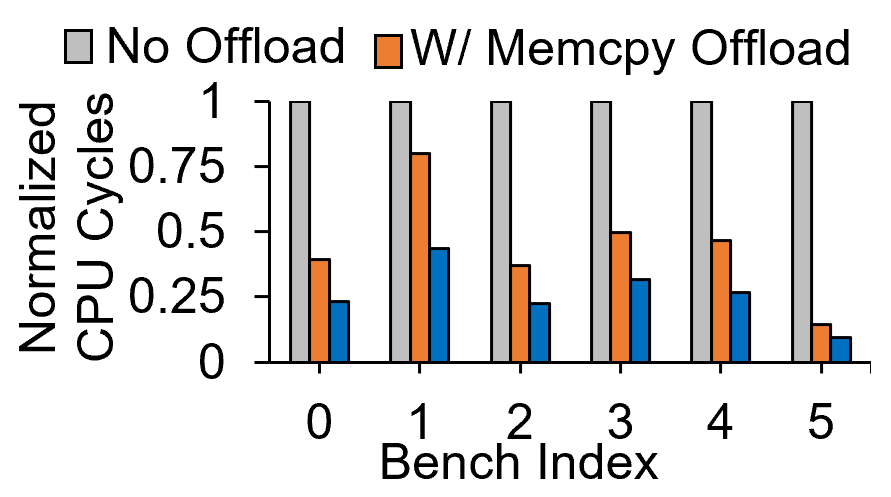}
    }
    \subfloat[DeathStarBench]{
        \label{fig:serialization_offload_deathstar}
        \includegraphics[width=0.48\linewidth]{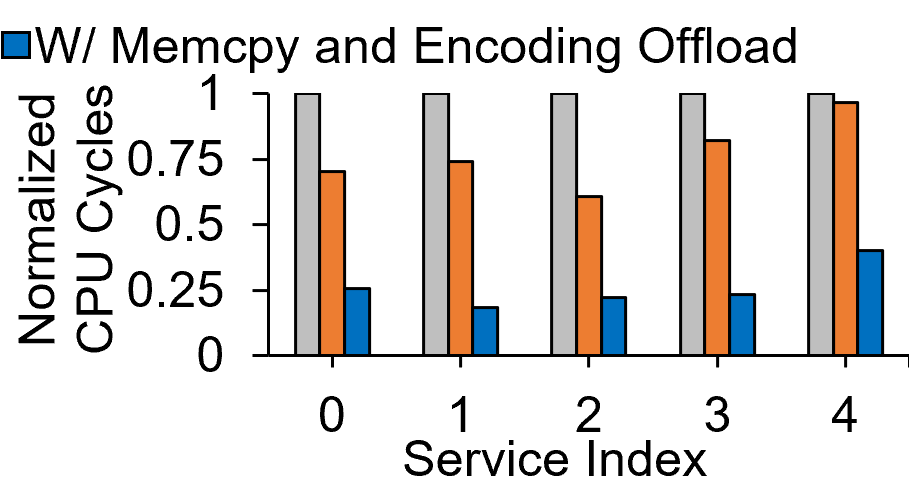}\
    } 
    \caption{Normalized CPU cycles of memory-affinity serializer with/without memcpy and encoding offload.} 
    \vspace{-3ex}
    \label{fig:encoding_offloading} 
\end{figure}

\vspace{0.2\baselineskip}
\noindent{\bf Effect of Encoding/memcpy Offload. }Figure~\ref{fig:encoding_offloading} demonstrates the normalized host CPU cycles with/without encoding/memcpy offloading for ``memory-affinity'' CPU-Accelerator co-serialization. The memcpy offload can reduce the host CPU cycles by an average of 55\% (23\%) in HyperProtoBench (DeathStarBench). Memcpy offload together with encoding offload can save the host CPU cycles by an average of 74\% (74\%) in HyperProtoBench (DeathStarBench). \sysname{} greatly reduces the host CPU cycles, indicating the effectiveness of ``memory-affinity'' CPU-Accelerator co-serialization. \cBlue{\sysname{} trades a few CPU cycles for the large decrease in the overall serialization time. The pre-serialization uses a geometric mean of 22\% CPU cycles compared with performing serialization in CPUs, while the geometric mean of the overall serialization time is decreased by 57\%.}


\label{evaluation:encoding_offloading}

\begin{figure}
	\centering
	{\includegraphics[keepaspectratio=true, width=\linewidth]{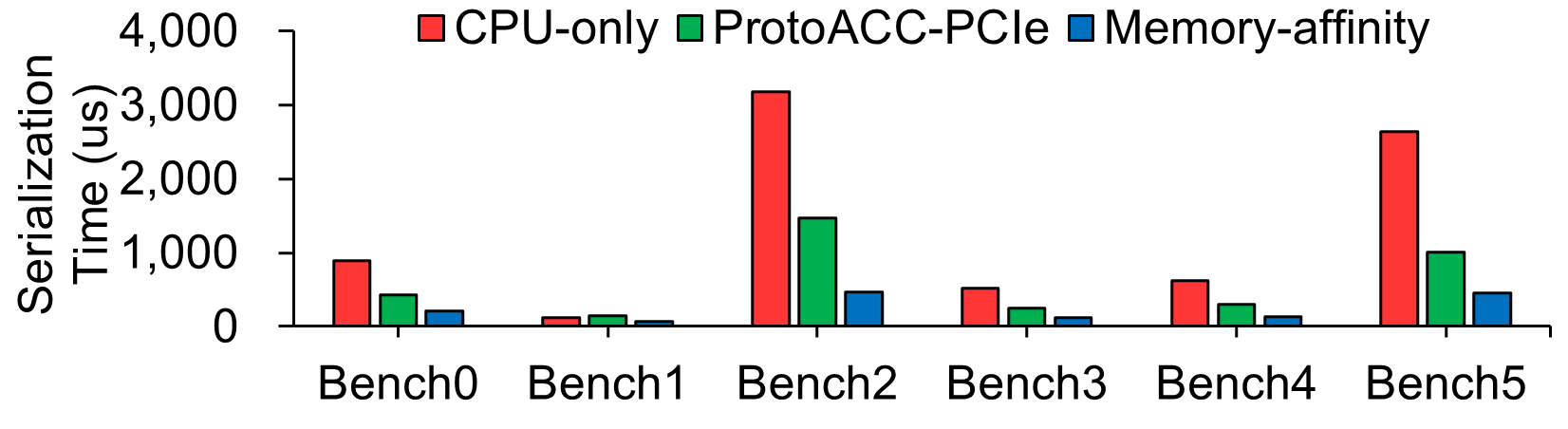}} 
        \vspace{-3ex}
	\caption{Serialization time comparisons.}
	\label{fig:serialization_time}
        \vspace{-3ex}
\end{figure}

\vspace{0.2\baselineskip}
\noindent{\bf Serialization Performance.} We then use HyperProtoBench to measure the end-to-end serialization time. 
We measure the serialization time of three design choices in Figure~\ref{fig:serialization_dataflow}.
As shown in Figure~\ref{fig:serialization_time}, ``Memory-affinity'' serialization spends the least time while ``CPU-only'' spends the most time. ``Memory-affinity'' outperforms ``ProtoACC-PCIe'' by 2.3$\times$ in geometric mean, because ``memory-affinity'' leverages the CPU or memcpy engine to perform pre-serialization for fields residing in the host CPU memory, instead through high-latency PCIe interconnect. ``Memory-affinity'' outperforms ``CPU-only'' by 4.3$\times$ in geometric mean, because pure software implementation of serialization incurs significant CPU overhead.

\begin{figure}[]
    \subfloat[Serialization]{
        \label{fig:bf3_serialization}
        \includegraphics[width=0.48\linewidth]{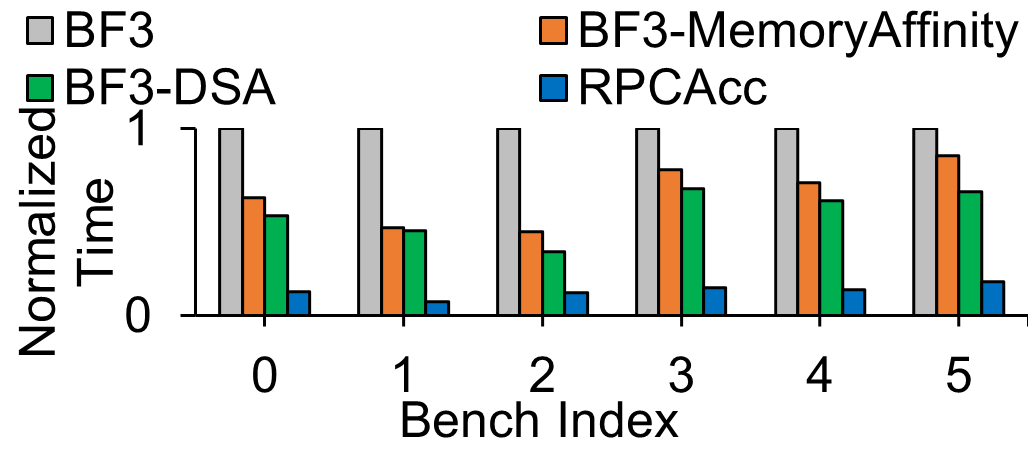}
    }
    \subfloat[Deserialization]{
        \label{fig:bf3_deserialization}
        \includegraphics[width=0.48\linewidth]{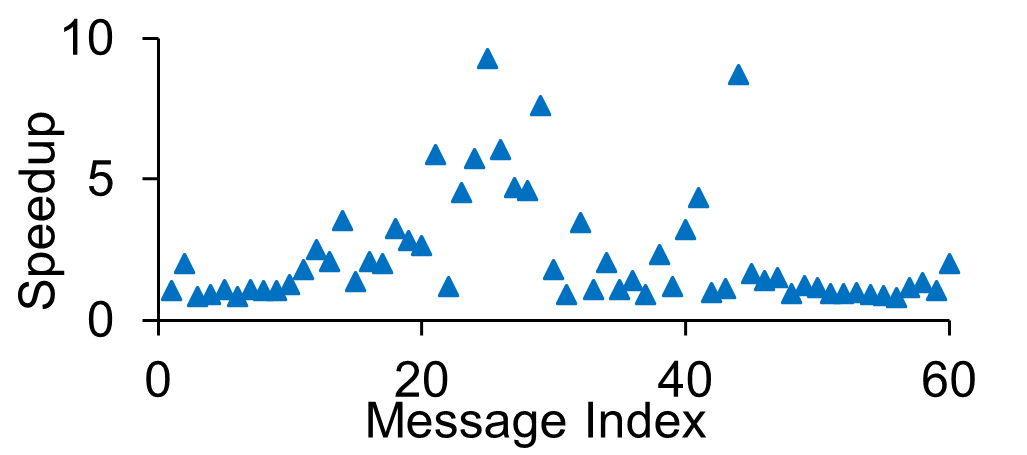}\
    } 
    \caption{Serialization time comparison and deserialization speedup of ``BF3-Oneshot'' over ``BF3''.} 
    \vspace{-3ex}
    \label{fig:bf3_ser_and_des} 
\end{figure} 

\vspace{-1ex}
\subsection{\cBlue{Comparison to SoC SmartNIC/DPU}}
\label{subsec:eval-soc}
\cBlue{In this section, we evaluate how \sysname{} optimizations can improve the performance when offloading RPC to a SoC-based SmartNIC, i.e., Nvidia Bluefield-3. ``BF3'' naively offloads the entire RPC to BF3, ``BF3-MemoryAffinity'' uses the host CPU to perform a pre-serialization process and uses BF3 cores to perform encoding/decoding. ``BF3-DSA'' is similar to ``BF3-MemoryAffinity'' and leverages the DSA memcpy engines during the host pre-serialization. ``BF3-Oneshot'' offloads the entire RPC to BF3 and coalesces small DMA requests into a large request during the deserialization.}

\cBlue{Figure~\ref{fig:bf3_serialization} shows the normalized serialization time on six benches of HyperProtoBench. We have three observations.  
First, applying pre-serialization optimization to the BF3 can reduce the serialization time by 1.58$\times$ on average. Second, applying memcpy offload optimization in the pre-serialization phase can additionally reduce the serialization time by 1.18$\times$ on average. The main reason for the speedup is that the pre-serialization can greatly reduce the high-latency cross-PCIe travesals and the memcpy offload can free host CPUs from copying large data fields. These results indicate applying \sysname{} optimizations to SoC-based SmartNIC/DPU can effectively reduce the overall serialization time. Third, \sysname{} still outperforms well-optimized BF3 implementation, this is mainly because \sysname{} uses hardware to perform CPU-inefficient encoding, while BF3 does this in the Arm cores. 
}

\cBlue{Figure~\ref{fig:bf3_deserialization} shows the deserialization throughput improvement of ``BF3-Oneshot'' over ``BF3''. Averagely applying one-shot DMA write optimization to BF3 can improve the deserialization throughput by 1.78$\times$, because one-shot DMA write can coalesce small DMA writes into a single large DMA write, improving PCIe transaction rate utilization. \sysname{} averagely achieves 5.9$\times$ higher deserialization throughput than ``BF3-Oneshot''. This is mainly because \sysname{} additionally offloads memory management and decoding to hardware.}

\cBlue{The above experiments indicate that \sysname{} optimizations can also works well on a SoC-based SmartNIC.}

\begin{figure}
	\centering
	{\includegraphics[keepaspectratio=true, width=\linewidth]{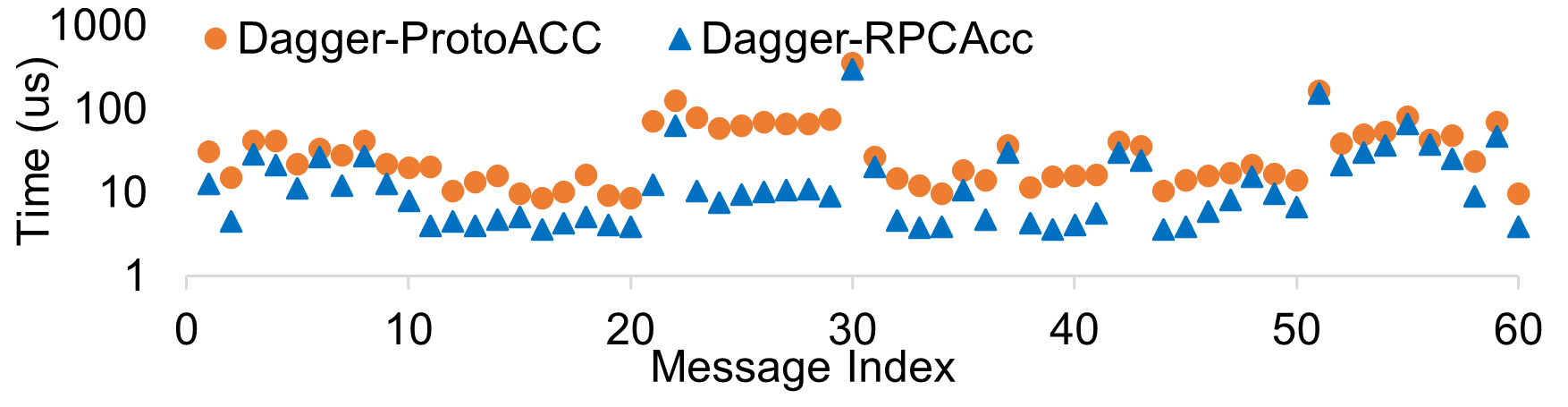}} 
        \vspace{-3ex}
	\caption{Serialization time of applying ProtoACC/\sysname{} serialization approach to Dagger.}
	\label{fig:dagger_serialization_time}
        \vspace{-3ex}
\end{figure}

\subsection{\cBlue{Comparison to Dagger}}
\label{subsec:eval-dagger}
\cBlue{Dagger does not support (de)serialization of structured and nested formats and naively adopting a hardware (de)serializer to Dagger would also suffer from long latency issue (FPGA's access over UPI incurs 400ns one-way latency, still much higher than CPU's memory access time). However, we can apply the optimizations proposed by \sysname{} to Dagger to improve RPC offloading performance for structured and nested RPC messages. We perform a cycle-accurate experiment that simulates integrating ProtoACC/\sysname{} (de)serialization methods into Dagger, called Dagger-ProtoACC and Dagger-\sysname{}, respectively. Both are clocked at 2 GHz. 
}



\cBlue{Figure~\ref{fig:dagger_serialization_time} shows the serialization time for two implementations. Dagger-\sysname{} averagely reduces the serialization time by 2.9$\times$, because applying \sysname{} serialization methods to Dagger can eliminate many cross-UPI traversals, where UPI has slightly lower latency than PCIe but still much higher than the normal memory access.}

\cBlue{For deserialization, \sysname{}’s one-shot DMA write mechanism can be easily adopted by Dagger to batch data writes within one RPC request, thus improving deserialization throughput at the cost of slightly increased deserialization latency. We do not present a throughput simulation, because how the PCIe/UPI transaction rate varies according to the data size is not known and we do not have real UPI-based hardware. However, Dagger paper has an inter-RPC batching mechanism and the authors claim that it can help improve data transfer efficiency. As such, we believe the intra-RPC batching (one-shot DMA write) can increase Dagger deserialization efficiency. 
Unlike throughput, the deserialization latency can be accurately simulated with one-way UPI latency (400ns) provided in the Dagger paper. Our evaluation on HyperProtoBench shows that adopting the one-shot DMA will only slightly increase the deserialization latency (geometric mean latency increases by 1.048x), which is acceptable considering the potential throughput benefits. }

\begin{figure}[]
    \subfloat[RX time.]{
        \label{fig:e_onchip_deserialization}
        \includegraphics[width=0.49\linewidth]{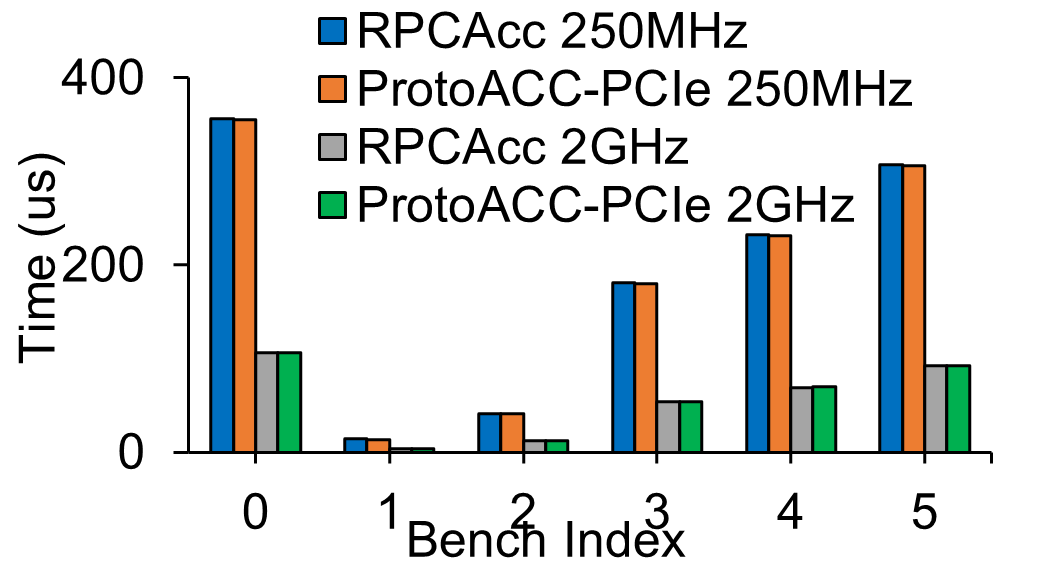}
    }
    \subfloat[TX time.]{
        \label{fig:e_onchip_serialization}
        \includegraphics[width=0.49\linewidth]{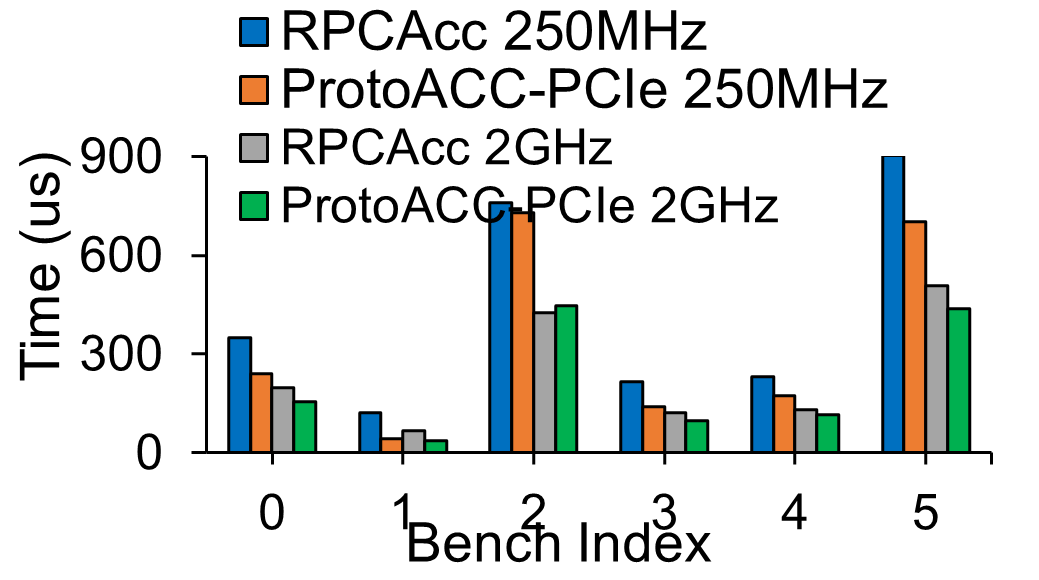}    
    }
    \caption{Time spent on the RPC layer.} 
    \vspace{-4ex}
    \label{fig:comparison_with_on_chip_acc} 
\end{figure} 

\subsection{Comparison to On-chip Accelerator}
\label{evaluation:comparison_with_on_chip_acc}
In this section, we compare the de/serialization time for \sysname{} and ``ProtoACC-OnChip'', using the HyperProtoBench. \sysname{} runs on our hardware platform while `ProtoACC-OnChip'' reports Xilinx Vivado simulation results as there is no real hardware. \cBlue{For \sysname{} at 250 Mhz, we report the measured RPC TX/RX time measured in the real hardware. The TX time is measured from when the CPU issues send RPC command to when the serialized data enters the NIC transport layer. The RX time is measured from when the data leaves the NIC transport layer to when the deserialized data arrives at the host CPU memory. For \sysname{} at 2 GHz, we first simulate the time spent on the accelerator. Then we manually add a PCIe transfer time and the time spent on the host CPU, both of which are measured when running 250 MHz real \sysname{} hardware.
For the on-chip accelerator simulation result, we first measure the simulated time spent on the RPC layer. Since the on-chip accelerator does not sit in the NIC, we then add an extra traversal time between the NIC and CPU memory.}

Figure~\ref{fig:comparison_with_on_chip_acc} shows the time consumed in the RPC layer, in receiving path (RX) and transmitting path (TX), respectively. We have two observations. 

First, when clocked at the same frequency, \sysname{} barely increases RX time compared to ``ProtoACC-OnChip''. This is because deserialization does not require pointer-chasing memory access, as such our PCIe-based deserializer would not incur performance degradation compared with a low-latency on-chip deserializer accelerator. 

\cBlue{Second, given that PCIe's latency (1250ns) is 17.9$\times$ higher than the memory latency (70ns)  setting of ``ProtoACC-OnChip'', ``ProtoACC-OnChip'' only achieves 1.4$\times$/1.24$\times$ lower TX time on average over \sysname{} when clocked at 250 MHz/2 GHz, respectively. This is mainly because \sysname{} enables the CPU pre-serialization to trade fast CPU memory copies for slow PCIe access.} 
What's more, when offloading an RPC kernel in \sysname{}, skipping accelerator fields can effectively reduce the TX time, further narrowing the serialization time gap. 


In sum, \sysname{} enables a PCIe-attached accelerator to achieve nearly the same deserialization performance and close serialization performance, compared to an on-chip de/serialization accelerator.

\begin{figure}[t]
    \subfloat[CU is preempted by other apps.]{
        \label{fig:e_remove_cu}
        \includegraphics[width=0.49\linewidth]{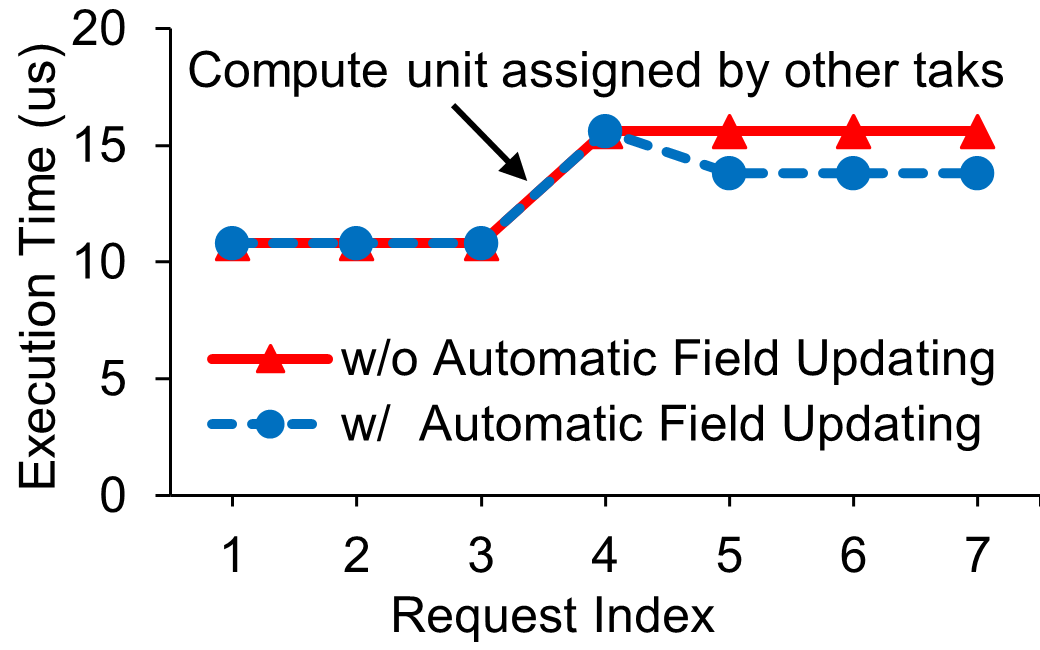}     
    }
    \subfloat[CU is reprogrammed by compression.]{
        \label{fig:e_add_cu}
        \includegraphics[width=0.49\linewidth]{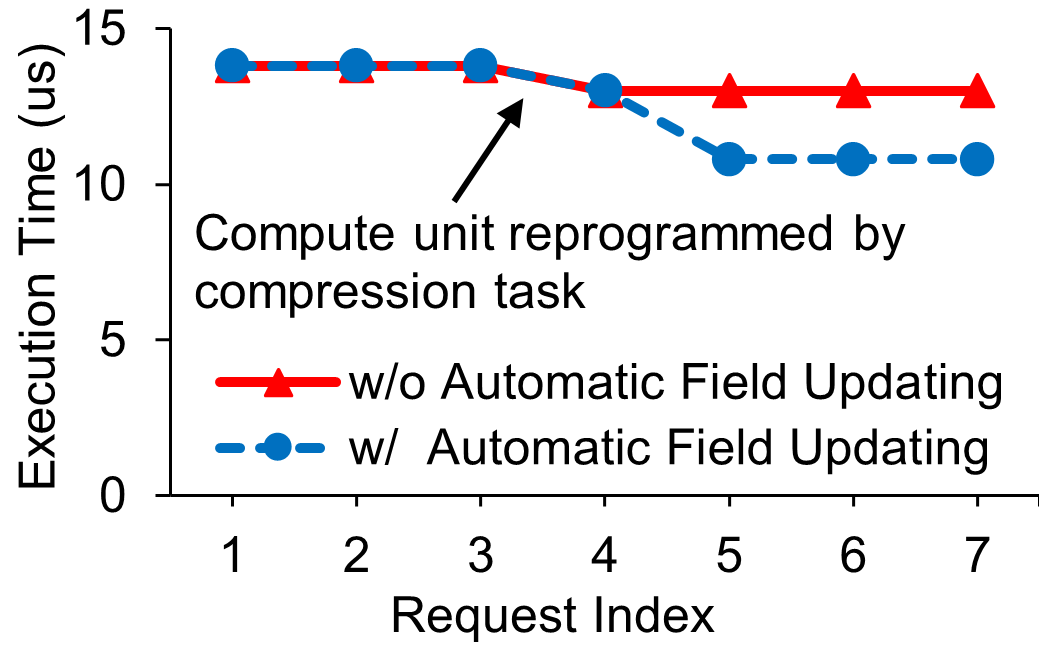}
    }
    \caption{Per-RPC execution time under kernel reconfiguration for the image compression example.} 
    \vspace{-3ex}
    \label{fig:e_automatic_field_updating} 
\end{figure} 

\subsection{Automatic Field Updating}
\label{subsec:eval-auto}
In this section, we evaluate automatic field updating using the image compression example in Listing~\ref{listing:example}. The accelerator is configured with one compression compute unit, which is reconfigured
to an unavailable state at runtime. It simulates the scenario when other applications have preempted the compute unit. When the compute unit is unavailable, the compression would switch to host CPU execution.

At the beginning of Figure~\ref{fig:e_remove_cu}, the large data field is put at accelerator memory and the compression is originally performed at the accelerator. After the 3rd request finishes, we set the compute unit as unavailable for the image compression service. The 4th request would suffer high execution latency since the large ``image'' field is put at the accelerator memory after serialization, CPU software has to manually move this field to CPU memory before performing compression on the CPU. Without automatic field updating, the execution time would remain high. With automatic field updating, all following requests' execution time would drop by several microseconds, because the explicit movement of the ``image'' field would update the schema table in the accelerator. This lets the deserializer put this field into the CPU memory next time, avoiding the CPU's explicit memory movement. 

Similarly, Figure~\ref{fig:e_add_cu} shows the situation that the CU is unavailable at the beginning and is available until the 4th request. With automatic field updating, the deserialization module can adapt to the dynamic change of compute units and put the field in the correct memory (CPU or accelerator memory). Besides, it eliminates the need for users to manually specify where fields should be placed after deserialization, saving a lot of hassle. As such, the automatic field updating mechanism yields high programmability. 

\subsection{End-to-end Application Performance}
\label{subsec:eval-end}

\begin{figure}[]
    \subfloat[Throughput.]{
        \label{fig:compression_tp}
        \includegraphics[width=0.48\linewidth]{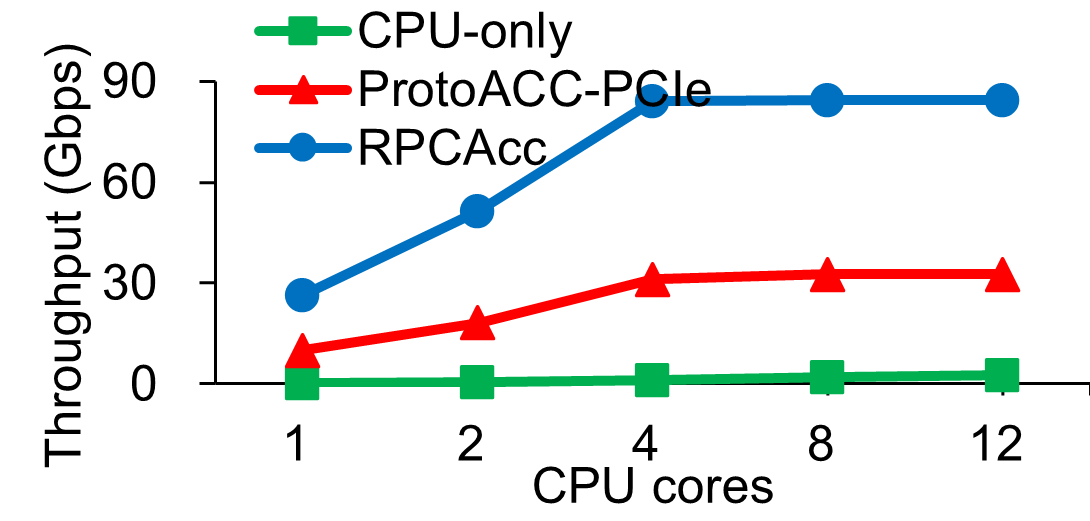}     
    }
    \subfloat[Average latency.]{
        \label{fig:compression_latency_average}
        \includegraphics[width=0.48\linewidth]{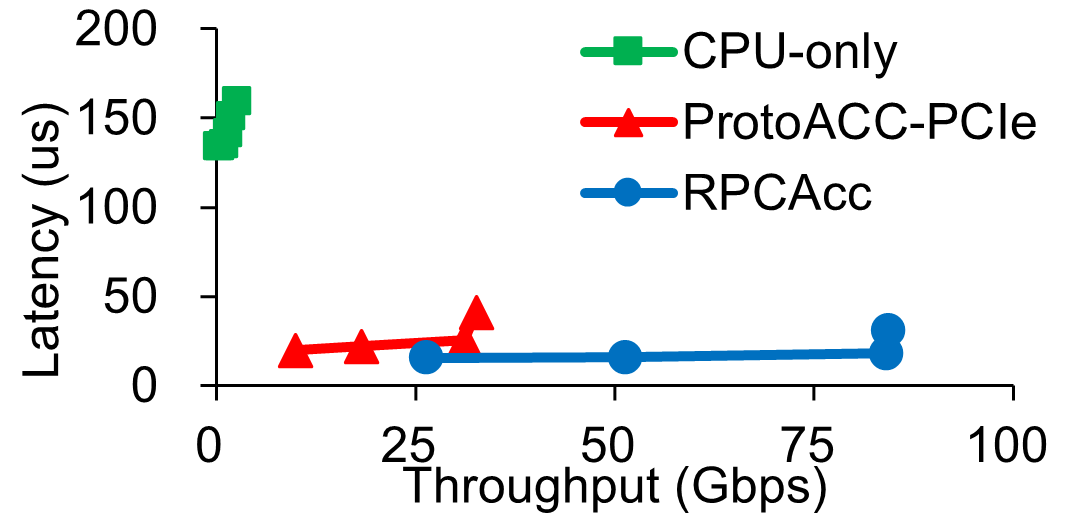}
    }
    \caption{Performanc comparison of three approaches when running an RPC-based image compression service.} 
    \vspace{-3ex}
    \label{fig:e_compression} 
\end{figure}

We evaluate \sysname{} using a practical cloud workload, which provides a high-performance and secure compression service.
Our workload mainly comprises three tasks for each RPC request: request authorization, compute-intensive compression, and encryption/decryption. For the ``CPU-only'' baseline, all three tasks run on the host CPU. For ``ProtoACC-PCIe'' and \sysname{}, compression and encryption/decryption run on the accelerator hardware, while the request authorization task is still conducted on the host CPU. The request authorization task is not offloaded because it usually only involves lightweight computation and changes frequently
Both ``ProtoACC-PCIe'' and \sysname{} are able to process the offloaded tasks at  line-rate (100 Gbps).

Figure~\ref{fig:compression_tp} shows the achieved throughput using the different numbers of host CPU cores. We observe that \sysname{}'s achievable throughput is 2.6$\times$ higher than the ``ProtoACC-PCIe'' baseline and 31.8 $\times$ higher than the ``CPU-only'' baseline. ``CPU-only'' performs worst because running compute-intensive compression, encryption/decryption, and RPC stack in the software is very inefficient compared with offloading them to hardware. \sysname{} outperforms ``ProtoACC-PCIe'' mainly because \sysname{} effectively offloads the RPC stack to a PCIe-attached accelerator with the proposed schemes. \cBlue{We also observe that skipping accelerator fields in the host CPU pre-serialization can save 65\% CPU cycles. This is because \sysname{} allows the KB-level large field to always reside in the accelerator memory and the hardware is responsible for its serialization.}


Figures~\ref{fig:compression_latency_average} shows the average latency of three implementations. We observe that \sysname{} can achieve 2.6 $\times$ (9.6 $\times$) lower average latency than the ``ProtoACC-PCIe'' (``CPU-only'') solution under the same throughput. \sysname{} outperforms the ``ProtoACC-PCIe'' baseline mainly because target-aware deserialization avoids much redundant data movement and memory-affinity serialization can greatly reduce the serialization time over the high-latency PCIe interconnect. 

\begin{figure}
	\centering
	{\includegraphics[keepaspectratio=true, width=0.8\linewidth]{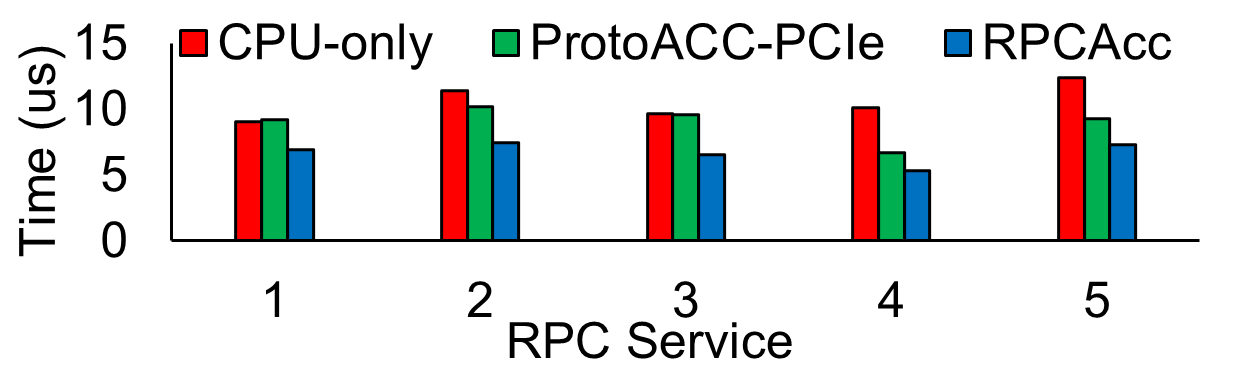}} 
        \vspace{-2ex}
	\caption{End-to-end execution time.}
	\label{fig:deathstar_e2e_time}
        \vspace{-2ex}
\end{figure}

\cBlue{To prove that \sysname{} can fit small-size RPCs, we perform an end-to-end comparison of five representative services (UniqueId, User, UrlShorten, SocialGraph, and ComposePost) in the widely-used DeathStarBench microservice suit. Figure~\ref{fig:deathstar_e2e_time} shows the end-to-end execution time. We observe that the geometric mean execution time of \sysname{} is 1.57$\times$/1.34$\times$ lower than the software baseline ``CPU-only''/``ProtoACC-PCIe''. This indicates that \sysname{} can also accelerate RPCs with a small message size.}

\subsection{Hardware Resource Usage}
\label{subsec:eval-hw}

Table~\ref{tab:resource} shows the FPGA resource consumption of ``ProtoACC-PCIe'' and \sysname{}. The resources of the offloaded RPC kernel are not reported as it is application-specific. \sysname{} is resource-frugal thanks to the compacted hardware data structures (schema table and temp buffer) and the streamlined serialization and serialization process. 


\begin{table} [t]
	\centering
	\begin{small}
 	\caption{FPGA resource consumption. }
	\label{tab:resource}	
	\vspace{-1.5ex}
	\begin{tabular}{|c||c|c|c|}
		\hline
		\textbf{Name} & \textbf{LUTs (K)} & \textbf{REGs (K)} & \textbf{BRAMs}  \\ 
		\hline
		\hline
            \textbf{ProtoACC-PCIe} & 221 (17\%) & 237(9\%) & 592 (29\%)\\
		\hline
            \hline
		\textbf{\sysname{}} & 170 (13\%) & 207 (8\%) & 552 (27\%)\\
            \hline
  
            

            {Serializer} & 30 (2.2\%) & 11 (0.41\%) & 55 (2.7\%)\\
            \hline

            {Deserializer} & 15 (1.1\%) & 7.8 (0.3\%) & 6 (0.3\%)\\

		\hline
	\end{tabular}
	\end{small}
        \vspace{-3ex}
\end{table}

\vspace{-1ex}
\section{Discussion}

\noindent\cBlue{{\bf How to accomodate differnt/evloving formats.} Currently, \sysname{} focuses on the widely used Protobuf format. But it is relatively easy to extend \sysname{} to support other formats such as Thrift~\cite{thrift}. In the following, we present two modifications. From the hardware perspective, we mainly need to modify the deserializer/serializer module and add transformation logic for Thrift fields. From the software perspective, we need to modify the compiler, adding parsing logic for ``.thrift'' files (which is similar to ``.proto'' files in Protobuf). }

\noindent\cBlue{{\bf (De)serialization-free formats.} Formats~\cite{flatbuffer, capnproto} such as Cap’n Proto~\cite{capnproto} are proposed to avoid the (de)serialization overheads during runtime. These formats sacrifice object mutability. However, the RPC message size is usually not known when it is created, so they need to allocate a large fixed buffer for each message which wastes memory space. Besides, the zero elements or unset fields would still occupy the space in the wire, incurring a larger transfer size than traditional formats like protobuf. To avoid these limitations, these formats also need additional designs to balance the (de)serialization overhead and waste of memory space/transfer size. Take Cap’n Proto as an example, to avoid memory waste, it allows a message to be split across multiple non-contiguous memory segments. Users can dynamically allocate more segments and use inter-segment pointers to link these segments together. To reduce the transfer size, Cap’n Proto adopts an operation called packing to compress these zero bytes in the wire format in serialization and unpack these zeros in deserialization. The packing/unpacking operations involve many bit-wise/byte-wise operations, and these CPU-inefficient operations incur similar overheads to that of encoding/decoding in traditional formats. We believe \sysname{}’s optimizations can mitigate the inefficiency introduced by the multiple segments and CPU-inefficient packing/unpacking. During serialization, we could add a CPU pre-process that copies incontiguous segments into a contiguous buffer, and the copy of large segments can be offloaded to CPU's on-chip memcpy engines, while the packing can be later executed in the NIC hardware. During deserialization, we could refer to the key idea of \sysname{}’s one-shot DMA write and decoding offloading. The unpacking operations can be offloaded to the NIC hardware and the DMA writes of different segments of one RPC message can be batched together to improve DMA transfer efficiency. In summary, it is easy to generalize \sysname{} to other formats. 
}

\noindent\cBlue{{\bf CXL.} 
RPCAcc’s idea still works well on top of a coherent fabric like CXL. Although the CXL coherence allows the host to access the accelerator memory using load/store (or reversely), the latency is still several times that of local memory access. We believe the key ideas of \sysname{} still hold: 1) putting the deserialized fields accordingly and letting CPU/Accelerator access their local memory as much as possible during the RPC process; 2) doing intra-RPC batching during the deserialization to improve transfer efficiency; 3) letting CPU/Accelerator serialize the fields in their local memory; 4) offloading bit-wise/byte-wise decoding/encoding and memory management to hardware. A coherent fabric like CXL will enhance \sysname{} in two main aspects. First, we can replace the costly MMIO-based mechanism with coherent memory access to implement the CPU-NIC interface. As such, the transaction rate for small RPC requests would not be bottlenecked by the low MMIO throughput. Second, it can avoid explicit cross-PCIe data movement at runtime. Our current implementation has to move the field explicitly, when the deserialized field is not in proper memory (CPU memory or accelerator memory). With CXL, both the CPU and accelerator can access the memory of each other using load/store instructions. 
CXL will enhance \sysname{} in two main aspects. First, we can replace the costly MMIO-based mechanism with coherent memory access to implement the CPU-NIC interface. As such, the transaction rate for small RPC requests would not be bottlenecked by low MMIO throughput. Second, it can avoid explicit cross-PCIe data movement at runtime. In the current implementation, when the deserialized field is not in proper memory (CPU memory or accelerator memory), the users have to move the field explicitly. With CXL, both the CPU and accelerator can access the memory of each other using load/store instructions. } 

\vspace{-2ex}
\section{Related Works}
\noindent{\bf Software-based RPC Acceleration.} 
Cornflakes~\cite{cornflakes_sosp23} leverages the scatter-gather capability to let NIC directly read the non-contiguous data from the host memory during the serialization. However, it requires that the data resides in the pinned DMA-safe region, which greatly harms memory utilization, especially in the cloud scenario. Besides, its serialization format does not contain encoding and is not compatible with existing applications.

\noindent{\bf Hardware-based RPC Acceleration.}
Prior works~\cite{cereal_isca20, dagger_asplos21, optimus_asplos20, cerebros_micro21, protoacc_micro21} offload the RPC stack or de/serialization to hardware to alleviate the CPU pressure. Cereal~\cite{cereal_isca20} adopts a special memory access interface to provide low-latency host memory access for the accelerator. Optimus Prime~\cite{optimus_asplos20} and Cerebros~\cite{cerebros_micro21} place an on-chip accelerator for de/serialization (entire RPC stack). ProtoACC~\cite{protoacc_micro21} proposes a novel near-core hardware accelerator for Protobuf. 
In contrast, \sysname{} focuses on optimizing RPC in a PCIe-attached accelerator, which is much more widely used in the modern cloud.

\vspace{0.2\baselineskip}
\noindent{\bf New Network Architecture.}
The nanoPU~\cite{nanopu_osdi21} is a new NIC-CPU co-design for RPC acceleration. It adds a fast path from the NIC directly to the CPU register file to achieve ultra-low latency packet access ($\sim$70ns). RAMBDA~\cite{rambda_hpca23} uses RDMA NICs and a standalone cache-coherent accelerator to accelerate data center applications. \sysname{} focus on the accelerations of RPC and de/serialization. 
NetDIMM~\cite{netdimm_micro19} integrates a full-blown NIC into the buffer device of a DIMM for fast data remote data access. FlexDriver~\cite{flexdriver_asplos22} allows the accelerator to control the NIC execution directly for high scalability.

\vspace{0.2\baselineskip}
\noindent{\bf Offloading to SmartNIC.} Many prior works~\cite{dpu_SC24, benchmarking_smartnic_osdi23, dpu_direct_TC24, Cowbird_sigcomm23, clicknp_sigcomm16,accelnet_nsdi18,hxdp,panic_osdi20,flowblaze_nsdi19,fpganic_atc22,smartds_isca23,turbo_hpca23,nica_atc19,bonola2022faster,clio_asplos22,floem_osdi18,gimbal_sigcomm21,ipipe_sigcomm19,e3_atc19,fairnic_sigcomm20,lambda_nic_icdcs20,fairnic_sigcomm20,xenic_asplos21,lognic_micro23,leed_sigcomm23} offload host tasks to FPGA or SoC SmartNICs to alleviate the host CPU pressure. None of these works tackle the problem of RPC tax. In contrast, \sysname{} offloads the RPC stack and computing kernels.

\vspace{0.2\baselineskip}
\noindent{\bf Header-payload Data Split.} Researchers have studied the header-payload split extensively~\cite{splitrpc_pomacs23,cachedirector_eurosys19,idio_micro22,nicmem_asplos22,nfslicer,p4_parking_conext20}. IDIO~\cite{idio_micro22} selectively disables Direct Cache Access (DCA) for the payload of received packets while always keeping DCA enabled for packet headers. SplitRPC~\cite{splitrpc_pomacs23} splits data using a fixed offset without de/serialization. In contrast, \sysname{} splits deserialized RPC messages according to RPC's fields and forwards them to either host CPU or Accelerator memory.

\vspace{-1ex}
\section{Conclusion}
This paper presents~\sysname{}, a hardware-software co-designed accelerator for reconfigurable RPC offloading. To tackle the ramifications of introducing PCIe, \sysname{} introduces three techniques: a target-aware deserializer, a memory-affinity CPU-accelerator collaborative serializer, and a runtime automatic field updating scheme. \sysname{} is an immediately deployed solution and provides the software abstraction to load the RPC stack and compute kernels on demand. 

\noindent{\bf Acknowledgement. }The work is supported by the following grants: 
the National Key R\&D Program of China (Grant No. 2022ZD0119301), 
the National Natural Science Foundation of China under the grant number (62472384), the Fundamental Research Funds for the Central Universities.

\bibliographystyle{plain}
\bibliography{content/references}

\end{document}